\documentclass[graybox]{svmult}

\usepackage{type1cm}        
\usepackage{makeidx}         
\usepackage{graphicx}        
\usepackage{multicol}        
\usepackage[bottom]{footmisc}

\usepackage{newtxtext}       %
\usepackage[varvw]{newtxmath}       

\makeindex             

\usepackage{listings}
\definecolor{darkgreen}{rgb}{0,0.6,0}
\lstdefinestyle{Python}{
    showstringspaces=false,
    language        = Python,
    basicstyle      = \small\ttfamily,
    morekeywords = {as},
    keywordstyle    = \color{blue},
    stringstyle     = \color{darkgreen},
    commentstyle    = \color{darkgreen}\ttfamily,
	breaklines = true,
	postbreak=\text{$\hookrightarrow$\space},
	alsoletter = {>,.} ,
    morekeywords = [2]{>>>,...},
    keywordstyle = [2]\color{cyan}\bfseries}

\allowdisplaybreaks

\begin{document}

\title*{Quasi-Monte Carlo Software}
\authorrunning{S.-C.\ T.\ Choi et al.}
\author{Sou-Cheng T. Choi \and Fred J. Hickernell \and R. Jagadeeswaran \and Michael J. McCourt \and Aleksei G. Sorokin}
\institute{Sou-Cheng T. Choi \at Department of Applied Mathematics, Illinois Institute of Technology,\\ RE 220, 10 W.\ 32$^{\text{nd}}$ St., Chicago, IL 60616; and Kamakura Corporation, 2222 Kalakaua Ave, Suite 1400, Honolulu, HI 96815 \email{schoi32@iit.edu}
\and
Fred J. Hickernell \at Center for Interdisciplinary Scientific Computation and \\
Department of Applied Mathematics, Illinois Institute of Technology \\ RE 220, 10 W.\ 32$^{\text{nd}}$ St., Chicago, IL 60616 \email{hickernell@iit.edu}
\and
R. Jagadeeswaran \at
Department of Applied Mathematics, Illinois Institute of Technology,\\ RE 220, 10 W.\ 32$^{\text{nd}}$ St., Chicago, IL 60616 \email{jrathin1@iit.edu}; and \\Wi-Tronix LLC, 631 E Boughton Rd, Suite 240, Bolingbrook, IL 60440
\and
Michael J. McCourt \at SigOpt, an Intel company, \\
100 Bush St., Suite 1100, San Francisco, CA 94104
\email{mccourt@sigopt.com}
\and 
Aleksei G. Sorokin \at
Department of Applied Mathematics, Illinois Institute of Technology,\\ RE 220, 10 W.\ 32$^{\text{nd}}$ St., Chicago, IL 60616 \email{asorokin@hawk.iit.edu}}

\maketitle

\abstract{Practitioners wishing to experience the efficiency gains from using low discrepancy sequences need correct, robust, well-written software.  This article, based on our MCQMC 2020 tutorial, describes some of the better quasi-Monte Carlo (QMC) software available.  We highlight the key software components required  by QMC to approximate multivariate integrals or expectations of functions of vector random variables.  We have combined these components in QMCPy, a Python open-source library, which we hope will draw the support of the QMC community.  Here we introduce QMCPy.}

\section{Introduction} \label{sec:intro} 

Quasi-Monte Carlo (QMC) methods promise great efficiency gains over independent and identically distributed (IID) Monte Carlo (MC) methods.  In some cases, QMC  achieves one hundredth of the error of IID MC in the same amount of time (see Figure \ref{fig:sc_comp}). Often, these efficiency gains are obtained simply by replacing IID sampling with low discrepancy (LD) sampling, which is the heart of QMC. 

Practitioners might wish to test whether QMC would speed up their computation.  Access to the best QMC algorithms available would make that easier.  Theoreticians or algorithm developers might want to demonstrate their ideas on various use cases to show their practical value.  

This tutorial points to some of the best QMC software available.  Then we describe  QMCPy \cite{QMCPy2020a}\footnote{QMCPy is in active development. This article is based on version 1.2 on PyPI.}, which is crafted to be a community-owned Python library that combines the best QMC algorithms and interesting use cases from various authors under a common user interface.

The model problem for QMC is approximating a multivariate integral,
\begin{equation} \label{eq:integral}
	\mu := \int_\mathcal{T} g(\boldsymbol{t}) \, \lambda(\boldsymbol{t}) \, \D \boldsymbol{t},
\end{equation}
where $g$ is the integrand, and $\lambda$ is a non-negative weight.  If $\lambda$ is a probability distribution (PDF) for the random variable $\boldsymbol{T}$, then $\mu$ is the mean of $g(\boldsymbol{T})$.  Regardless, we perform a suitable variable transformation to interpret this integral as the  mean of a function of a multivariate, standard uniform random variable:
\begin{equation} \label{eq:fintegral}
	\mu = \mathbb{E}[f(\boldsymbol{X})] =  \int_{[0,1]^d}  f(\boldsymbol{x}) \,  \D \boldsymbol{x} , \qquad \boldsymbol{X} \sim \mathcal{U}[0,1]^d.
\end{equation}

QMC approximates the population mean, $\mu$,  by a sample mean,
\begin{equation} \label{eq:samplemean}
	\widehat{\mu} := \frac 1n \sum_{i=0}^{n-1} f(\boldsymbol{X}_i), \qquad \boldsymbol{X}_0, \boldsymbol{X}_1, \ldots \overset{\textup{M}}{\sim} \mathcal{U}[0,1]^d.
\end{equation}
The choice of the sequence $\{\boldsymbol{X}_i\}_{i=0}^\infty$ and the choice of $n$ to satisfy  the prescribed error requirement,
\begin{equation} \label{eq:err_req}
	\lvert\mu - \widehat{\mu}\rvert \le \varepsilon \qquad \text{absolutely or with high probability},
\end{equation} 
are important decisions, which  QMC software helps the user make.

Here, the notation $\overset{\textup{M}}{\sim}$ means that the sequence mimics the specified, target distribution, but not necessarily in a probabilistic way.  We  use this notation in two forms:  $\overset{\textup{IID}}{\sim}$ and $\overset{\textup{LD}}{\sim}$.

\begin{figure}[t]
	\includegraphics[height=5.8cm]{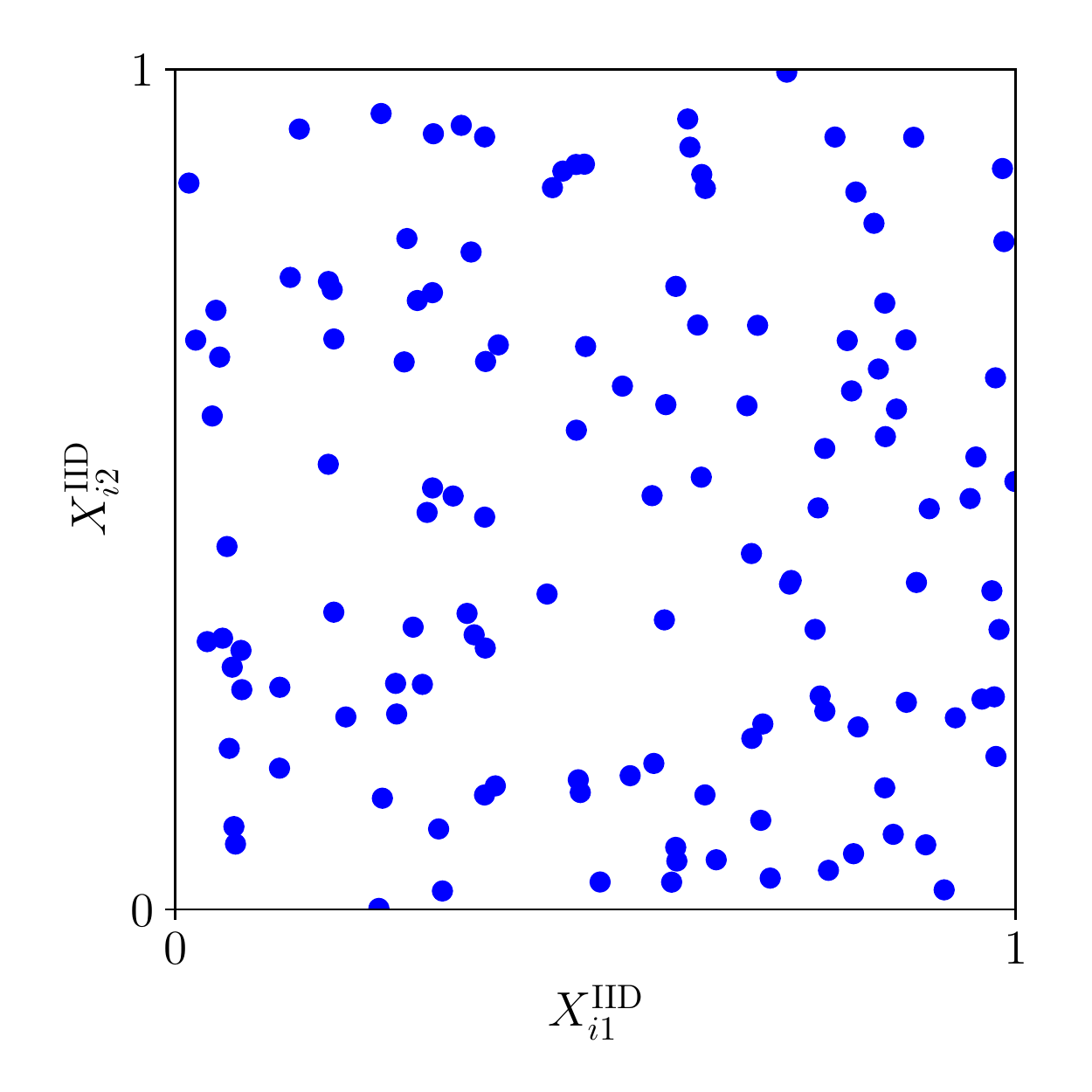}
	\quad
	\includegraphics[height=5.8cm]{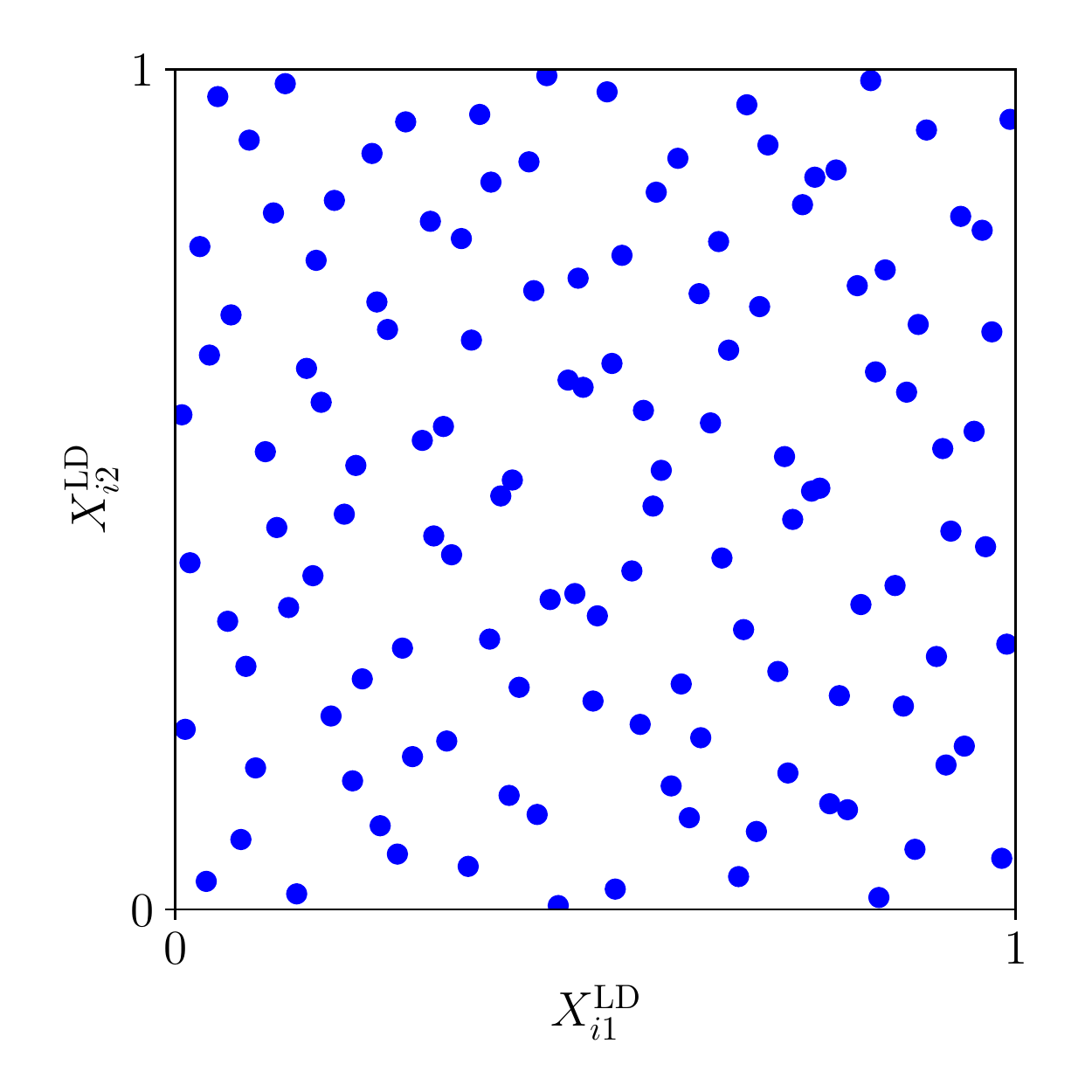}
	\caption{IID points (left) contrasted with LD points (right).  The LD points cover the square more evenly.}
	\label{fig:comparePts}
\end{figure}

IID sequences must be random. The position of any point is not influenced by any other, so clusters and gaps occur.  A randomly chosen subsequence of an IID sequence is also IID.  When we say that $\boldsymbol{X}_0, \boldsymbol{X}_1, \ldots \overset{\textup{IID}}{\sim} F$ for some distribution $F$, we mean that for any positive integer $n$, the  multivariate probability distribution of $\boldsymbol{X}_0, \ldots, \boldsymbol{X}_{n-1}$ is the product of the marginals, specifically,
\begin{equation*}
	F_{n}(\boldsymbol{x}_0, \ldots, \boldsymbol{x}_{n-1}) = F(\boldsymbol{x}_0) \cdots  F(\boldsymbol{x}_{n-1}).
\end{equation*}
When IID points are used to approximate $\mu$ by the sample mean, the root mean squared error is $\mathcal{O}(n^{-1/2})$.  Fig.~\ref{fig:comparePts} displays IID uniform points, $\boldsymbol{X}^{\textup{IID}}_0, \boldsymbol{X}^{\textup{IID}}_1, \ldots \overset{\textup{IID}}{\sim} \mathcal{U}[0,1]^2$, i.e.,  the target distribution is $F_{\textup{unif}}(\boldsymbol{x}) = x_1 x_2$.

LD sequences may be deterministic or random, but each point is carefully coordinated with the others so that they fill the domain well.  Subsequences of LD sequences are generally not LD.  When we say that $\boldsymbol{X}_0, \boldsymbol{X}_1, \ldots \overset{\textup{LD}}{\sim} \mathcal{U}[0,1]^d$, we mean that for any positive integer $n$,  the \emph{empirical distribution} of $\boldsymbol{X}_0, \ldots, \boldsymbol{X}_{n-1}$, denoted $F_{\{\boldsymbol{X}_i\}_{i=0}^{n-1}}$,  approximates the uniform distribution, $F_{\textup{unif}}$, well (relative to $n$).  (The empirical distribution of a set assigns equal probability to each point.)

A measure of the difference between the empirical distribution of a set of points and the uniform distribution is called a \emph{discrepancy} and is denoted $D(\{\boldsymbol{X}_i\}_{i=0}^{n-1})$ \cite{DicEtal14a,Hic97a,Hic99a,Nie92}.  This is the origin of the term ``low discrepancy'' points or sequences.  LD points by definition have a smaller discrepancy than IID points.  Fig.~\ref{fig:comparePts} contrasts IID uniform points with LD points, $\boldsymbol{X}^{\textup{LD}}_0, \boldsymbol{X}^{\textup{LD}}_1 \ldots \overset{\textup{LD}}{\sim} \mathcal{U}[0,1]^2$, in this case, linearly scrambled and digitally shifted Sobol' points.

The error in using the sample mean to approximate the integral can be bounded according to the Koksma-Hlawka inequality and its extensions \cite{DicEtal14a,Hic97a,Hic99a,Nie92} as the product of the discrepancy of the sampling sequence and the variation of the integrand, denoted $V(\cdot)$:
\begin{equation}
	\lvert\mu - \widehat{\mu}\rvert = \biggl \lvert \int_{[0,1]^d} f(\boldsymbol{x}) \, \D \bigl(F_{\textup{unif}} - F_{\{\boldsymbol{X}_i\}_{i=0}^{n-1}}\bigr) (\boldsymbol{x}) \biggr \rvert \le D\bigl(\{\boldsymbol{X}_i\}_{i=0}^{n-1}\bigr) V(f).
\end{equation} 
The variation is a (semi-) norm of the integrand in a suitable Banach space.  The discrepancy corresponds to the norm of the error functional for that Banach space.  For typical Banach spaces, the discrepancy of LD points is $\mathcal{O}(n^{-1+\epsilon})$, a higher convergence order than for IID points.  For details,  readers may refer to the references.

Here, we expect the reader to see in Fig.~\ref{fig:comparePts} that the LD points cover the integration domain more evenly than IID points.  LD sampling can be thought of as a more even distribution of the sampling sites than IID.  LD sampling is similar to stratified sampling.  In the examples below, the reader will see the demonstrably smaller cubature errors arising from using LD points.

In the sections that follow, we first overview available QMC software.  We next describe an architecture for good QMC software, i.e., what the key components are and how they should interact.  We then describe how we have implemented this architecture in QMCPy.  Finally, we summarize further directions that we hope QMCPy and other QMC software projects will take.  Those interested in following or contributing to the development of QMCPy are urged to visit the GitHub repository at https://github.com/QMCSoftware/QMCSoftware.


We have endeavored to be as accurate as possible at the time of writing this article.  We hope that progress in QMC software development will make this article happily obsolete in the coming years.

\section{Available Software for QMC} \label{sec:available} 
QMC software spans  LD sequence generators, cubatures, and applications.  Here we review the better-known software, recognizing that some software overlaps multiple categories. Whenever applicable, we state each library's accessibility in QMCPy, or contrast its functionalities with QMCPy's --- where we lag behind in QMCPy, we strive to catch up in the near future.

Software focusing on generating high-quality LD sequences and their generators includes the following, listed in alphabetical order:
\begin{description}
	\item[\textbf{BRODA}] \sloppypar Commerical and non-commercial software developed jointly with I.M.~Sobol' in C++, MATLAB, and Excel~\cite{BRODA20a}. BRODA can generate Sobol' sequences up to 65,536 dimensions. In comparison, QMCPy supports Sobol' sequences up to 21,201 dimensions.	
	\item[\textbf{Burkhardt}] Various QMC software for generating van der Corput, Faure, Halton, Hammersley,  Niederreiter, or Sobol' sequences in C, C++, Fortran, MATLAB, or Python~\cite{Bur20a}. In QMCPy, we have implemented digital net, lattice, and Halton generators.
	\item[\textbf{LatNet Builder}] The successor to \emph{Lattice Builder}~\cite{LEcMun14a}, this is a C++ library with Python and Java interfaces (in SSJ below) for generating vectors or matrices for lattices and digital nets~\cite{LatNet,LEcEtal22a}. QMCPy contains a module for parsing the resultant vectors or matrices from LatNet Builder for compatibility with our LD point generators. 
	\item[\textbf{MATLAB}] Commercial software for scientific computing \cite{MAT9.10}, which contains Sobol' and Halton sequences in the Statistics and Machine Learning Toolbox. Both generators can be applied jointly with the Parallel Computing Toolbox to accelerate their execution speed. The  dimension of the Sobol' sequences is restricted to 1,111, which is relatively small, yet sufficient for most applications. 
	\item[\textbf{MPS}] Magic Point Shop contains lattices and Sobol' sequences in C++, Python, and MATLAB~\cite{Nuy17a}.   QMCPy started with MPS for developing LD generators.
	\item[\textbf{Owen}] Owen's randomized Halton sequences with dimensions up to 1,000 \cite{Owe20a} and scrambled Sobol' sequences with dimensions up to 21,021  \cite{owen1998scrambling, owen2021r}  in R. QMCPy  supports Owen's Halton randomization method, and we plan to implement Owen's nested uniform scrambling for digital nets in the near future.
	\item[\textbf{PyTorch}] Open-source Python library for deep learning, with unscrambled or scrambled Sobol' sequences \cite{NEURIPS2019_9015,PyTorch}. PyTorch enables seamless utilization of Graphics Processing Units (GPUs) or Field Programmable Gate Arrays (FPGAs).
	\item[\textbf{QMC.jl}] LD Sequences in Julia \cite{Rob20a}. Julia~\cite{bezanson2012julia} is an interpreted language similar to Python and R in terms of ease of use, but is designed to run much faster.
	\item[\textbf{qrng}] Randomized Sobol', Halton, and Korobov sequences in R~\cite{QRNG2020}. The default Halton randomization in QMCPy utilizes the methods from qrng. 
	\item[\textbf{SciPy}] Scientific computing library in Python with Latin hypercube,  Halton, and Sobol' generators \cite{2020SciPy-NMeth}.
	\item[\textbf{TF Quant Finance}] Google's Tensorflow  deep-learning library~\cite{tensorflow2015-whitepaper} specialized for financial modeling \cite{tfqf21}. It contains lattice and Sobol' generators alongside with algorithms sped up with GPUs, FGPAs, or Tensor Processing Units (TPUs).
\end{description}
Software focusing on QMC cubatures and applications includes the following:
\begin{description}
	\item[\textbf{GAIL}] The Guaranteed Automatic Integration Library  contains automatic (Q)MC stopping criteria in MATLAB~\cite{ChoEtal21a,HCJJ14}. These are iterative procedures for one- or high-dimensional integration that take a user's input error tolerance(s) and determine the number of (Q)MC sampling points necessary to achieve user-desired accuracy (almost surely).  Most of GAIL's (Q)MC functions, some with enhancements, are implemented in Python in QMCPy.
	\item[\textbf{ML(Q)MC}] Multi-Level (Q)MC routines in C, C++, MATLAB, Python, and R~\cite{GilesSoft}. We have ported ML(Q)MC functions to QMCPy.
	\item[\textbf{MultilevelEstimators.jl}] ML(Q)MC methods in Julia \cite{Rob21}. The author, Pieterjan Robbe, has contributed cubature algorithms and use cases to QMCPy.
	\item[\textbf{OpenTURNS}] Open source initiative for the Treatment of Uncertainties, Risks'N Statistics \cite{OpenTURNS} written in C++ and Python, leveraging R statistical packages, as well as LAPACK and BLAS for numerical linear algebra.
	\item[\textbf{QMC4PDE}] \sloppypar QMC for elliptic PDEs with random diffusion coefficients in Python~\cite{KuoNuy16a}.
	\item[\textbf{SSJ}] Stochastic Simulation with the \texttt{hups} package in Java \cite{SSJ}.
	\item[\textbf{UQLab}] Framework for Uncertainty Quantification in MATLAB \cite{UQLab2014}.  The core of UQLab is closed source, but a large portion of the library is open source. Recently,  UQ[py]Lab, the beta release of UQLab with Python bindings, is  available as Software as a Service (SaaS) via UQCloud~\cite{lataniotis2021uncertainty}.

\end{description}

The sections that follow describe QMCPy \cite{QMCPy2020a}, which is our attempt to establish a framework for QMC software and to combine the best of the above software under a common user interface written in Python 3.  The choice of language was determined by the desire to make QMC software accessible to a broad audience, especially the technology industry.

\section{Components of QMC Software}
QMC cubature can be summarized as follows.  We want to approximate the expectation, $\mu$, well by the sample mean, $\widehat{\mu}$, where \eqref{eq:integral}, \eqref{eq:fintegral}, and \eqref{eq:samplemean} combine to give
\begin{multline} \label{eq:cubSummary}
	\mu : = \int_\mathcal{T} g(\boldsymbol{t}) \, \lambda(\boldsymbol{t}) \, \D \boldsymbol{t}  = \mathbb{E}[f(\boldsymbol{X})] = \int_{[0,1]^d} f(\boldsymbol{x}) \, \D \boldsymbol{x} \approx \frac 1n \sum_{i=0}^{n-1} f(\boldsymbol{X}_i) =: \widehat{\mu}, \\
	 \boldsymbol{X} \sim \mathcal{U}[0,1]^d, \ \boldsymbol{X}_0, \boldsymbol{X}_1, \ldots \overset{\textup{M}}{\sim} \mathcal{U}[0,1]^d.
\end{multline}
Moreover, we want to satisfy the error requirement in \eqref{eq:err_req}.
This requires four components, which we implement as QMCPy classes.

\begin{description}
	
	\item[\textbf{Discrete Distribution}]  produces the sequence $\boldsymbol{X}_0, \boldsymbol{X}_1, \dots$ that mimics $\mathcal{U}[0,1]^d$;
	
	\item[\textbf{True Measure}] $\boldsymbol{t} \mapsto \lambda (\boldsymbol{t}) \D \boldsymbol{t}$  defines the original measure, e.g., Gaussian or Lebesgue;
	
	\item[\textbf{Integrand}] $g$  defines the original integrand, and $f$ defines the transformed version to fit the \texttt{DiscreteDistribution}; and
	
	\item[\textbf{Stopping Criterion}] determines how large $n$ should be to ensure that $\lvert \mu - \widehat{\mu}\rvert \le \varepsilon$ as in (\ref{eq:err_req}).
\end{description}

The software libraries referenced in Section \ref{sec:available} provide one or more of these components. QMCPy combines multiple examples of all these components under an object-oriented framework. Each example is implemented as a concrete class that realizes the properties and methods required by the abstract class for that component. The following sections detail descriptions and specific examples for each component. 

Thorough documentation of all QMCPy classes is available in \cite{QMCPyDocs}. Demonstrations of how QMCPy works are given in Google Colab notebooks \cite{QMCPyTutColab2020,QMCPyTutColab2020_paper}. The project may be installed from PyPI into a Python 3 environment via the command \texttt{pip install qmcpy}. In the code snippets that follow, we assume QMCPy has been imported alongside NumPy \cite{numpy} via the following commands in a Python Console:
\begin{lstlisting}[style=Python]
>>> import qmcpy as qp
>>> import numpy as np
\end{lstlisting}

\section{Discrete Distributions}

LD sequences typically mimic $\mathcal{U}[0,1]^d$.  Good sequences mimicking other distributions are obtained by transformations as described in the next section.  We denote by \texttt{DiscreteDistribution} the abstract class containing LD sequence generators.  In most cases that have been implemented, the points $\boldsymbol{X}_0, \ldots, \boldsymbol{X}_{n-1}$ have an empirical (discrete) distribution that closely approximates the uniform distribution, say, in the sense of discrepancy.  We also envision future possible \texttt{DiscreteDistribution} objects that assign unequal weights to the sampling points.  In any case, the term ``discrete'' refers to the fact that these are sequences of points (and weights), not continuous distributions.

QMCPy implements \emph{extensible} LD sequences, i.e., those that allow practitioners to obtain and use $\boldsymbol{X}_n, \boldsymbol{X}_{n+1}, \ldots $ without discarding $\boldsymbol{X}_0, \ldots, \boldsymbol{X}_{n-1}$.  Halton sequences do not have preferred sample sizes $n$, but extensible integration lattices and digital sequences in base $b$ prefer $n$ to be a power of $b$.  For integration lattices and digital sequences, we have focused on base $b=2$ since this is a popular choice and for convenience in generating extensible sequences.

Integration lattices and digital sequences in base $2$ have an elegant group structure, which we summarize in Table \ref{tab:GroupProp}.  The addition operator  is $\oplus$\footnote{The operator $\oplus$ is commonly used to denote exclusive-or, does correspond to its meaning for digital sequences in base $2$.  However, we are using it here in a more general sense.}, and its inverse is $\ominus$.  The unshifted sequence is $\boldsymbol{Z}_0, \boldsymbol{Z}_1, \ldots$ and the randomly shifted sequence is $\boldsymbol{X}_0, \boldsymbol{X}_1, \ldots$

\begin{table}[h]
	\centering
	\caption{Properties of lattices and digital net sequences.  Note that they share group properties but also have distinctives.} \label{tab:GroupProp}
\[
	\renewcommand{\arraystretch}{1.3}
\begin{array}{c@{\qquad}c}
	\hline
	\multicolumn{2}{c}{\text{Define \ldots}} \\
	\multicolumn{2}{c}{\boldsymbol{Z}_1, \boldsymbol{Z}_2, \boldsymbol{Z}_4, \ldots \in [0,1)^d \text{ chosen well} } \\
	\multicolumn{2}{c}{
	\boldsymbol{Z}_{i} := i_0  \boldsymbol{Z}_1 \oplus i_1 \boldsymbol{Z}_{2} \oplus i_2  \boldsymbol{Z}_{4} \oplus  i_3  \boldsymbol{Z}_{8} \oplus \cdots 
	\quad
	\text{for }i = i_0 +i_1 2 + i_2 4 + i_3 8 + \cdots, \; i_\ell \in \{0,1\}} \\
    \multicolumn{2}{c}{\boldsymbol{X}_i := \boldsymbol{Z}_i \oplus \boldsymbol{\Delta}, \qquad \text{where }\boldsymbol{\Delta} \overset{\textup{IID}}{\sim} [0,1)^d} \\  \hline
	\text{Rank-1 Integration Lattices} & \text{Digital Nets} \\
		\boldsymbol{t} \oplus \boldsymbol{x} : = (\boldsymbol{t} + \boldsymbol{x}) \bmod \boldsymbol{1} & \boldsymbol{t} \oplus \boldsymbol{x} := \text{binary digitwise addition}, \oplus_{\textup{dig}} \\ 
		\text{require } \boldsymbol{Z}_{2^{m}} \oplus \boldsymbol{Z}_{2^{m}} = \boldsymbol{Z}_{\lfloor2^{m-1}\rfloor}
		\quad \forall m \in \mathbb{N}_0 
		\\[.5ex]
\hline
\multicolumn{2}{c}{\text{Then it follows that \ldots}} \\
	\multicolumn{2}{c}{\left . \begin{array}{r}
			\mathcal{P}_m := \{\boldsymbol{Z}_0, \ldots, \boldsymbol{Z}_{2^m-1}\}, \quad
			\boldsymbol{Z}_i \oplus \boldsymbol{Z}_j \in \mathcal{P}_m \\
			\mathcal{P}_{\boldsymbol{\Delta},m} := \{\boldsymbol{X}_0, \ldots, \boldsymbol{X}_{2^m-1}\}, \quad
			\boldsymbol{X}_i \oplus \boldsymbol{X}_j \ominus \boldsymbol{X}_k \in \mathcal{P}_{\boldsymbol{\Delta},m}
	\end{array} \right \}\quad \begin{array}{l}\forall  i,j,k \in \{0, \ldots, 2^{m} -1\} \\ \forall m \in \mathbb{N}_0\end{array}} \\
\hline
\end{array}
\]
\end{table}

We illustrate lattice and Sobol' sequences using QMCPy. First, we create an instance of a $d=2$ dimensional \texttt{Lattice} object of the  \texttt{DiscreteDistribution} abstract class. Then we generate the first eight (non-randomized) points in this lattice. 
\begin{lstlisting}[style=Python]
>>> lattice = qp.Lattice(dimension=2, randomize=False)
>>> lattice.gen_samples(n=8)
"ParameterWarning:
    Non-randomized lattice sequence includes the origin"
array([[0.   , 0.   ],
       [0.5  , 0.5  ],
       [0.25 , 0.75 ],
       [0.75 , 0.25 ],
       [0.125, 0.375],
       [0.625, 0.875],
       [0.375, 0.125],
       [0.875, 0.625]])
\end{lstlisting}
The first three generators for this lattice are $\boldsymbol{Z}_1 = (0.5, 0.5)$, $\boldsymbol{Z}_2 = (0.25, 0.75)$, and $\boldsymbol{Z}_4 = (0.125, 0.375)$.  One can check that $(\boldsymbol{Z}_2 + \boldsymbol{Z}_4) \bmod \boldsymbol{1} = (0.375, 0.125) = \boldsymbol{Z}_6$, as Table \ref{tab:GroupProp} specifies.

The random shift has been turned off above to illuminate the group structure.  We normally include the randomization to ensure that there are no points on the boundary of $[0,1]^d$.  Then, when points are transformed to mimic distributions such as the Gaussian, no LD points will be transformed to infinity.  Turning off the randomization generates a warning when the \texttt{gen\_samples} method is called.

Now, we generate Sobol' points using a similar process as we did for lattice points.  Sobol' sequences are one of the most popular example of digital sequences.
\begin{lstlisting}[style=Python]
>>> sobol = qp.Sobol(2, randomize=False)
>>> sobol.gen_samples(8, warn=False)
array([[0.   , 0.   ],
       [0.5  , 0.5  ],
       [0.25 , 0.75 ],
       [0.75 , 0.25 ],
       [0.125, 0.625],
       [0.625, 0.125],
       [0.375, 0.375],
       [0.875, 0.875]])
\end{lstlisting}
Here, $\boldsymbol{Z}_4$ differs from that for lattices, but more importantly, addition for digital sequences differs from that for lattices.  Using digitwise addition for digital sequences, we can confirm that according to Table \ref{tab:GroupProp},
\begin{align*}
\boldsymbol{Z}_2 \oplus_{\textup{dig}} \boldsymbol{Z}_4 = (0.25,0.75)  \oplus_{\textup{dig}} (0.125,0.625) \\
=  ({}_20.010,{}_20.110)  \oplus_{\textup{dig}} ({}_20.001,{}_20.101) &= ({}_20.011,{}_20.011) \\
& = (0.375,0.375) = \boldsymbol{Z}_6.
\end{align*}

By contrast, if we construct a digital sequence using the generators for the lattice above with $\boldsymbol{Z}_2 = (0.25, 0.75)$, and $\boldsymbol{Z}_4 = (0.125, 0.375)$, we would obtain
\begin{multline*}
\boldsymbol{Z}_6 = \boldsymbol{Z}_2 \oplus_{\textup{dig}} \boldsymbol{Z}_4   = ({}_20.010,{}_20.110)  \oplus_{\textup{dig}} ({}_20.001,{}_20.011)  \\
= ({}_20.011,{}_20.101) = (0.375, 0.625),
\end{multline*}
which differs from the $\boldsymbol{Z}_6=(0.375, 0.125)$ constructed for lattices.  To emphasize, lattices and digital sequences are different, even if they share the same generators, $\boldsymbol{Z}_1, \boldsymbol{Z}_2, \boldsymbol{Z}_4, \ldots$.

The examples of \texttt{qp.Lattice} and \texttt{qp.Sobol} illustrate how QMCPy LD generators share a common user interface.  The dimension is specified when the instance is constructed, and the number of points is specified when the \texttt{gen\_samples} method is called.  Following Python practice, parameters can be input without specifying their names if input in the prescribed order.  QMCPy also includes Halton sequences and IID sequences, again deferring details to the QMCPy documentation~\cite{QMCPyDocs}.

A crucial difference between IID generators and LD generators is reflected in the behavior of generating $n$ points.  For an IID generator, asking for $n$ points repeatedly gives different points each time because they are meant to be random and independent.
\begin{lstlisting}[style=Python]
>>> iid = qp.IIDStdUniform(2); iid.gen_samples(1)
array([[0.40538109, 0.12255759]])
>>> iid.gen_samples(1)
array([[0.50913741, 0.11312201]])
\end{lstlisting}
Your output may look different depending on the seed used to generate these random numbers.

On the other hand, for an LD generator, asking for $n$ points repeatedly gives \emph{the same} points each time because they are meant to be the first $n$ points of a specific LD sequence.  
\begin{lstlisting}[style=Python]
>>> lattice = qp.Lattice(2); lattice.gen_samples(1)
array([[0.2827584 , 0.36731649]])
>>> lattice.gen_samples(1)
array([[0.2827584 , 0.36731649]])
\end{lstlisting}
Here we allow the randomization so that the first point in the sequence is not the origin.  To obtain the \emph{next} $n$ points, one may specify the start and ending indices of the sequence.
\begin{lstlisting}[style=Python]
>>> lattice.gen_samples(2)
array([[0.2827584 , 0.36731649],
       [0.7827584 , 0.86731649]])
>>> lattice.gen_samples(n_min=1, n_max=2)
array([[0.7827584 , 0.86731649]])
\end{lstlisting}

\begin{figure}[t]
	\includegraphics[width=1\textwidth]{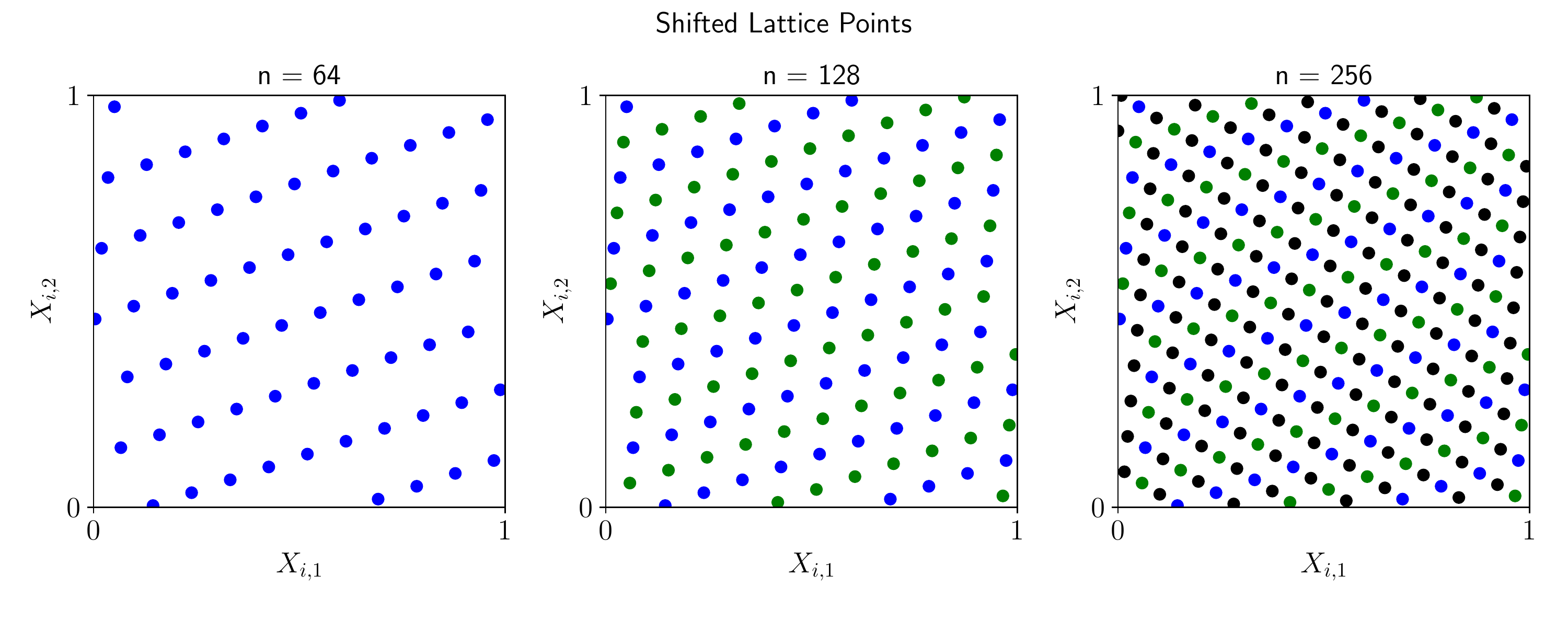}
	\qquad
	\includegraphics[width=1\textwidth]{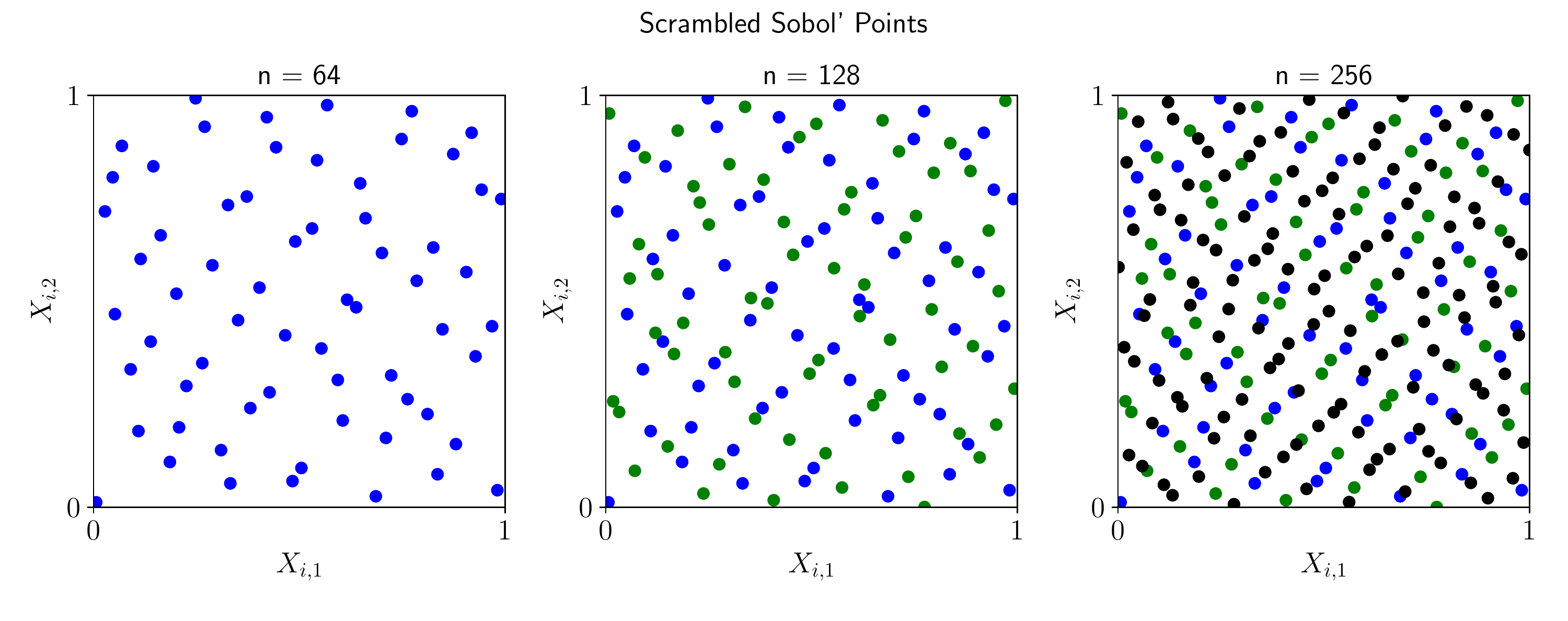}
	\caption{Randomized lattice and Sobol' points mimicking a $\mathcal{U}[0,1]^2$ measure for $n = 64, 128,$ and 256. Note how increasing the number of points evenly fills in the gaps between the points.}
	\label{fig:increase_n}
\end{figure}

Fig.~\ref{fig:increase_n} shows how increasing the number of lattice and Sobol' LD points through powers of two fills in the gaps in an even way.

\section{True Measures}

The LD sequences implemented as \texttt{DiscreteDistribution} objects mimic the $\mathcal{U}[0,1]^d$ distribution.  However, we may need sequences to mimic other distributions.  This is implemented via variable transformations, $\boldsymbol{\Psi}$.  In general, if $\boldsymbol{X} \overset{\textup{M}}{\sim} \mathcal{U}[0,1]^d$, then
\begin{subequations} \label{eq:exampleVarTrans}
\begin{gather}
\boldsymbol{T} = \boldsymbol{\Psi}(\boldsymbol{X}) := \boldsymbol{a}  + (\boldsymbol{b} - \boldsymbol{a}) \odot \boldsymbol{X} \overset{\textup{M}}{\sim}  \mathcal{U}[\boldsymbol{a},\boldsymbol{b}], \\
\label{eq:exampleVarTransGauss}
\boldsymbol{T} = \boldsymbol{\Psi}(\boldsymbol{X}) := \boldsymbol{a} + \mathsf{A} \boldsymbol{\Phi}^{-1}(\boldsymbol{X})  \overset{\textup{M}}{\sim} \mathcal{N}(\boldsymbol{a}, \Sigma), \\
\nonumber  \text{where }  \boldsymbol{\Phi}^{-1}(\boldsymbol{X}) : = \begin{pmatrix} \Phi^{-1}(X_1) \\ \vdots \\ \Phi^{-1}(X_d)\end{pmatrix}, \qquad \Sigma = \mathsf{A} \mathsf{A}^T,
\end{gather}
\end{subequations}
and $\odot$ denotes term-by-term (Hadamard) multiplication.  Here, $\boldsymbol{a}$ and $\boldsymbol{b}$ are assumed to be finite, and $\Phi$ is the standard Gaussian distribution function.  Again we use $\overset{\textup{M}}{\sim}$ to denote mimicry, not necessarily in a probabilistic sense.

\begin{figure}[t]
	\includegraphics[width=.45\textwidth]{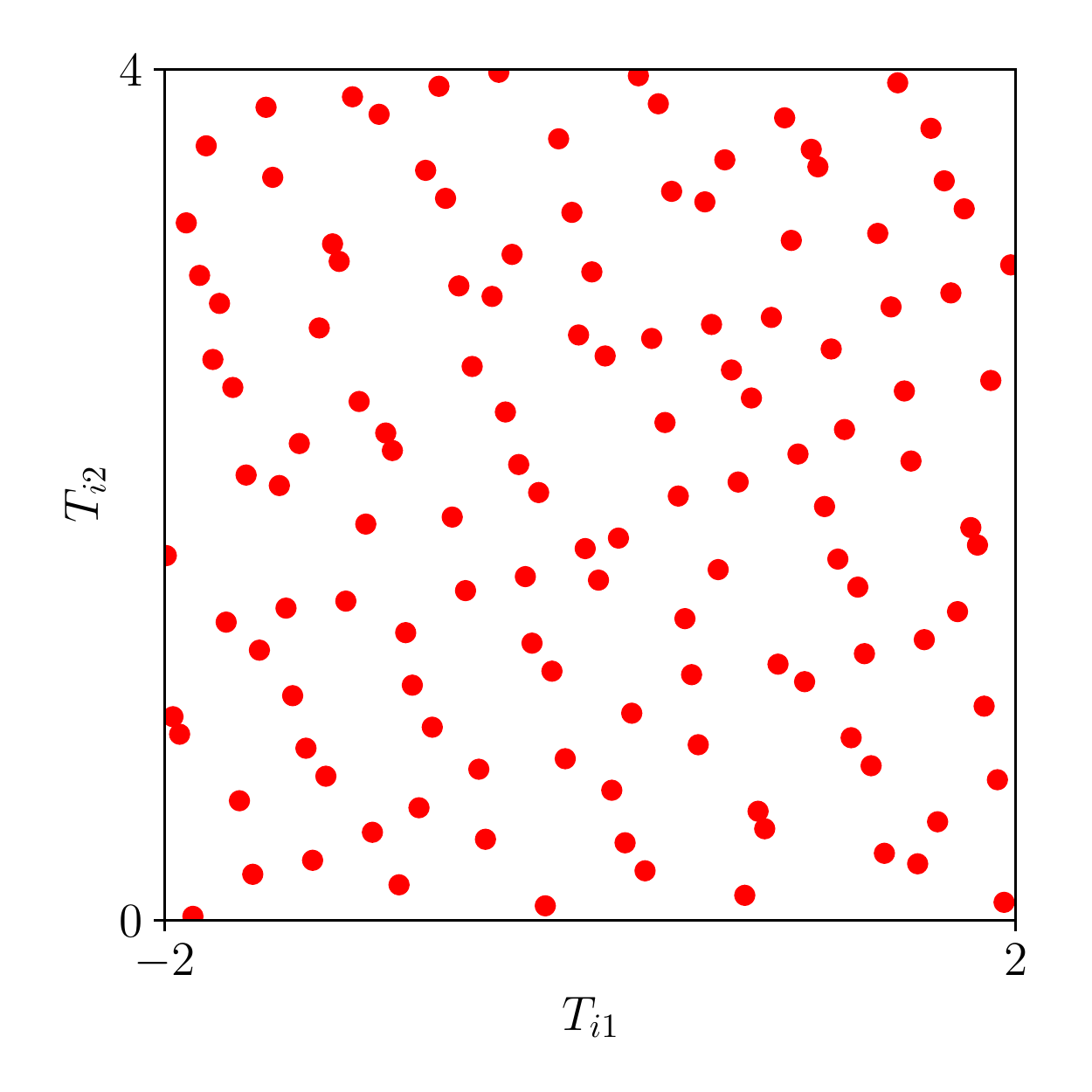} 
	\includegraphics[width=.45\textwidth]{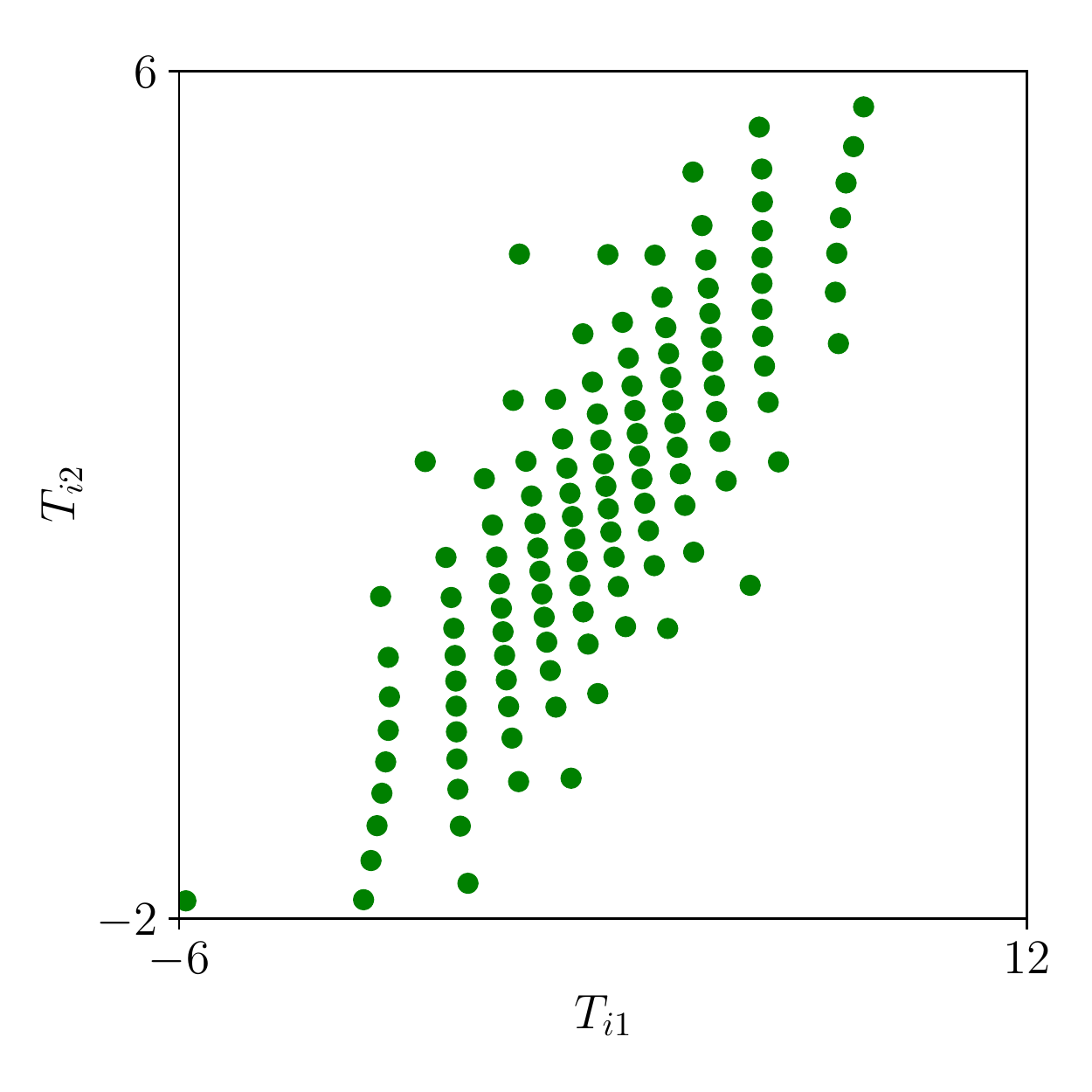}
	\caption{Halton samples transformed to mimic a uniform $\mathcal{U}\left( \begin{bmatrix} -2 \\ 0 \end{bmatrix},\begin{bmatrix} 2 \\ 4 \end{bmatrix} \right)$ 
	distribution (left) and lattice samples transformed to mimic a Gaussian $\mathcal{N}\left(\begin{bmatrix} 3 \\ 2 \end{bmatrix}, \begin{bmatrix} 9 & 5 \\ 5 & 4 \end{bmatrix} \right)$ distribution (right).}
	\label{fig:tm_ug}
\end{figure}

Fig.~\ref{fig:tm_ug} displays LD sequences transformed as described above to mimic a uniform and a Gaussian distribution.  The code to generate these points takes the following form for uniform points based on a Halton sequence: 
\begin{lstlisting}[style=Python]
>>> u = qp.Uniform(
...         sampler = qp.Halton(2),
...         lower_bound = [-2, 0],
...         upper_bound = [2, 4])
>>> u.gen_samples(4)
array([[ 1.80379772,  3.51293599],
       [-0.19620228,  0.84626932],
       [ 0.80379772,  2.17960265],
       [-1.19620228,  3.95738043]])
\end{lstlisting}
whereas for Gaussian points based on a lattice sequence, we have:
\begin{lstlisting}[style=Python]
>>> g = qp.Gaussian(qp.Lattice(2),
...         mean =        [3, 2], 
...         covariance = [[9, 5], 
...                       [5, 4]])
>>> g.gen_samples(4)
array([[-0.00920667,  0.29392389],
       [ 5.02422481,  1.45408341],
       [ 6.53688109,  5.14785917],
       [ 2.58403124,  1.16941969]])
\end{lstlisting}
Here the covariance decomposition $\Sigma = \mathsf{A} \mathsf{A}^T$ is done using principal component analysis. The Cholesky decomposition is also available.

\begin{figure}[t]
	\includegraphics[width=1\textwidth]{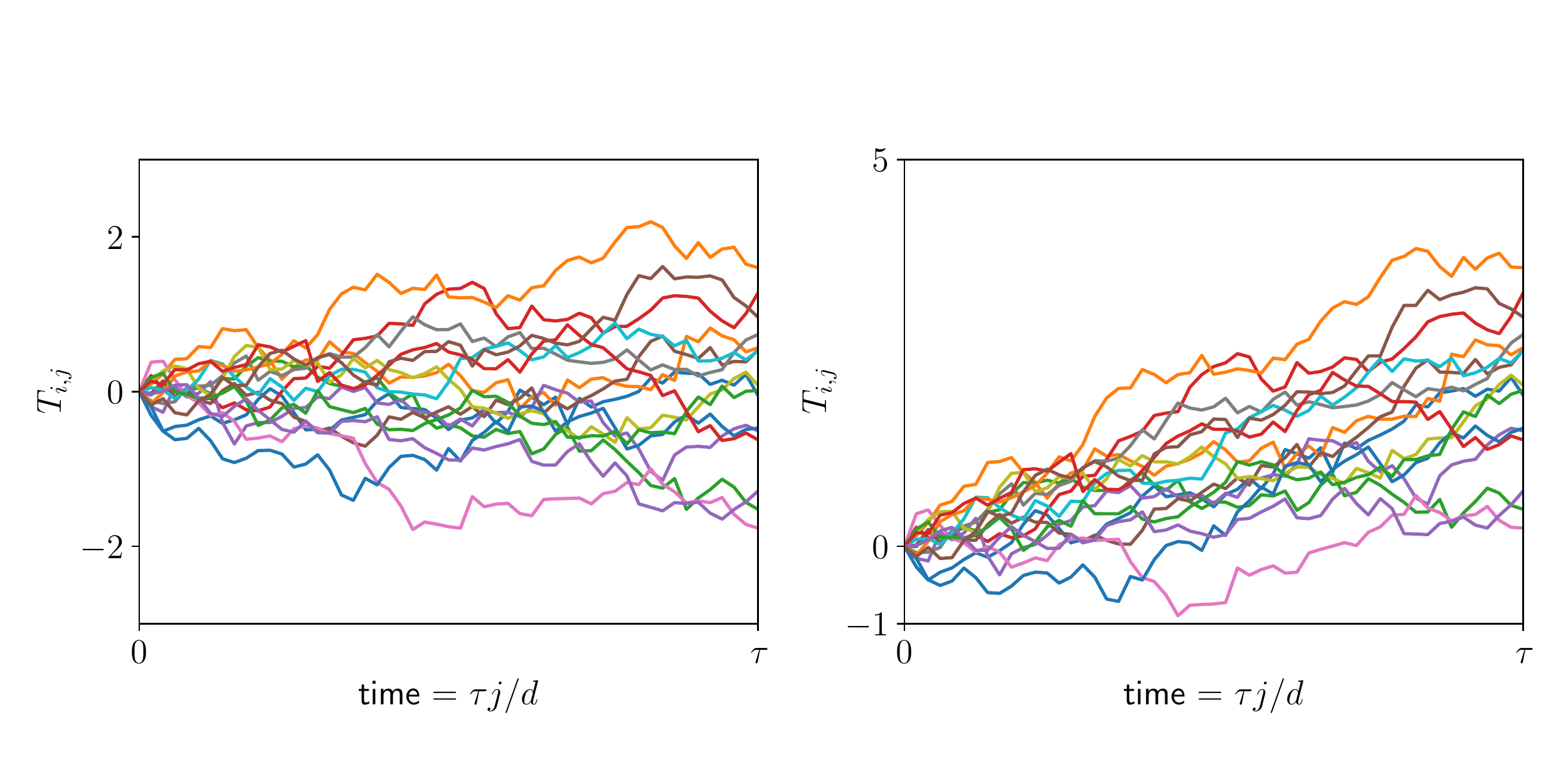} 
	\caption{Sobol' samples transformed to mimic a 52-dimensional Brownian Motion without drift (left) and with drift coefficient 2 (right).}
	\label{fig:tm_bm}
\end{figure}

The Brownian motion distribution arises often in financial risk applications.  Here the $d$ components of the variable $\boldsymbol{T}$ correspond to the discretized Brownian motion at times $\tau/d, 2\tau/d, \ldots, \tau$, where $\tau$ is the time horizon.  The distribution is a special case of the Gaussian with covariance 
\begin{equation} \label{eq:BMcov}
	\Sigma = (\tau/d) \bigl (\min(j,k) \bigr)_{j,k=1}^d
\end{equation}
and mean $\boldsymbol{a}$, which  is proportional to the times $(\tau/d)(1, 2, \ldots, d)^T$. The code for generating a Brownian motion is
\begin{lstlisting}[style=Python]
>>> bm = qp.BrownianMotion(qp.Sobol(4), drift=2)
>>> bm.gen_samples(2)
array([[-0.28902829,  0.48582198,  1.81113976,  2.2376372 ],
       [ 1.41552651,  1.93926124,  1.27619672,  1.47496128]])
\end{lstlisting}
Fig.~\ref{fig:tm_bm} displays a Brownian motion based on Sobol' sequence with and without a drift.

\section{Integrands}

Let's return  to the integration problem in \eqref{eq:integral}, which we must rewrite as \eqref{eq:fintegral}.  We choose a transformation of variables defined as $\boldsymbol{t} = \boldsymbol{\Psi}(\boldsymbol{x})$ where $\boldsymbol{\Psi}:[0,1]^d \to \mathcal{T}$.  This leads to 
\begin{align}
	\nonumber 
 \mu &= \int_\mathcal{T} g(\boldsymbol{t}) \, \lambda(\boldsymbol{t}) \, \D \boldsymbol{t}  = \int_{[0,1]^d} g\bigl(\boldsymbol{\Psi}(\boldsymbol{x})\bigr) \, \lambda\bigl(\boldsymbol{\Psi}(\boldsymbol{x})\bigr) \,\lvert \boldsymbol{\Psi}'(\boldsymbol{x})\rvert  \, \D \boldsymbol{x} =  \int_{[0,1]^d}f(\boldsymbol{x}) \, \D \boldsymbol{x},  \\
 \label{eq:transVar}
  & \qquad \qquad \text{where } f(\boldsymbol{x})  = g\bigl(\boldsymbol{\Psi}(\boldsymbol{x})\bigr)  \, \lambda\bigl(\boldsymbol{\Psi}(\boldsymbol{x})\bigr) \,\lvert \boldsymbol{\Psi}'(\boldsymbol{x})\rvert ,
\end{align}
and $\lvert \boldsymbol{\Psi}'(\boldsymbol{x})\rvert := \lvert \partial \boldsymbol{\Psi}/\partial \boldsymbol{x}\rvert$ represents the Jacobian of the variable transformation.  The abstract class \texttt{Integrand} provides $f$ based on the user's input of $g$ and the \texttt{TrueMeasure} instance, which defines $\lambda$ and the transformation $\boldsymbol{\Psi}$. Different choices of $\boldsymbol{\Psi}$ lead to different $f$, which may give different rates of convergence of the cubature, $\widehat{\mu}$ to $\mu$.

We illustrate the \texttt{Integrand} class via an example of Keister \cite{Kei96}:
\begin{equation} \label{eq:KeisterIntegral}
	\mu 
	= \int_{\mathbb{R}^d} \cos(\lVert \boldsymbol{t} \rVert) \exp(-\boldsymbol{t}^T \boldsymbol{t}) \, \D \boldsymbol{t} \\ 
	= \int_{\mathbb{R}^d} \underbrace{\pi^{d/2} \cos(\lVert \boldsymbol{t} \rVert)}_{g(\boldsymbol{t})}\, \underbrace{\pi^{-d/2} \exp(-\boldsymbol{t}^T \boldsymbol{t}) }_{\lambda(\boldsymbol{t})} \, \D \boldsymbol{t}.
\end{equation}
Since $\lambda$ is the density for $\mathcal{N}(\boldsymbol{0},\mathsf{I}/2)$, it is natural to choose $\boldsymbol{\Psi}$ according to \eqref{eq:exampleVarTransGauss} with $\mathsf{A} = \sqrt{1/2} \, \mathsf{I}$, in which case $\lambda(\boldsymbol{\Psi}(\boldsymbol{x})) \lvert \boldsymbol{\Psi}'(\boldsymbol{x})\rvert  = 1$, and so 
\[
\mu = \int_{[0,1]^d} \underbrace{\pi^{d/2} \cos(\lVert \boldsymbol{\Psi}(\boldsymbol{x}) \rVert)}_{f(\boldsymbol{x})} \, \D \boldsymbol{x}, \qquad 
\boldsymbol{\Psi}(\boldsymbol{x}) := \sqrt{1/2} \,\boldsymbol{\Phi}^{-1}(\boldsymbol{x}).
\]

\begin{figure}[t]
	\includegraphics[height=7cm]{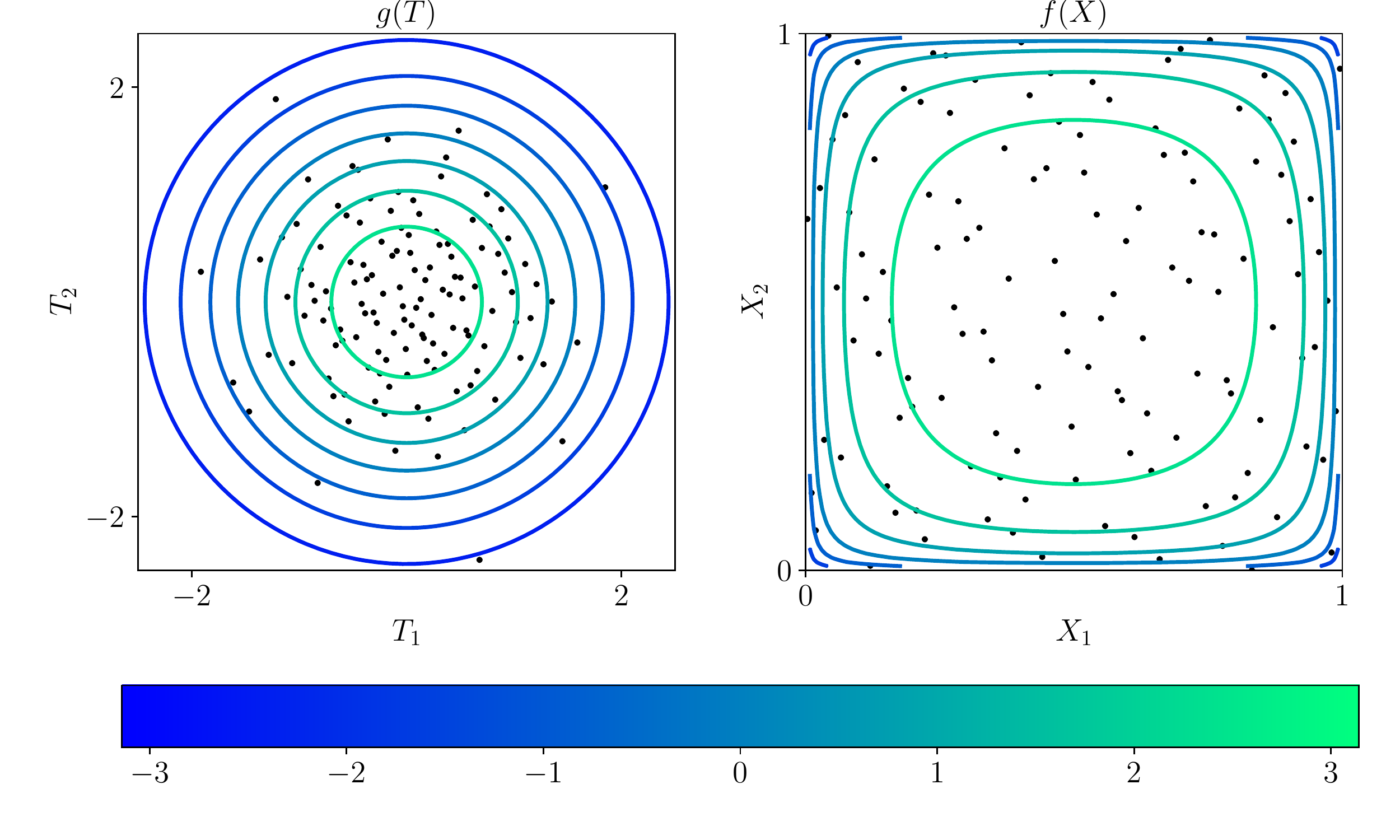}
	\caption{Right: Sampling the transformed Keister integrand $f$ at Halton points $\boldsymbol{X}_i \overset{\textup{LD}}{\sim} \mathcal{U}[0,1]^2$. Left: Sampling the original Keister integrand $g$ at $\boldsymbol{T}_i = \boldsymbol{\Psi}(\boldsymbol{X}_i) \overset{\textup{M}}{\sim} \mathcal{N}(\boldsymbol{0},\mathsf{I}/2)$ where $\boldsymbol{\Psi}$ is defined in \eqref{eq:exampleVarTransGauss}.  } \label{fig:ikc}
\end{figure}

The code below sets up an \texttt{Integrand} instance using QMCPy's \texttt{CustomFun} wrapper to tie a user-defined function $g$ into the QMCPy framework.  Then we evaluate the sample mean of $n=1000$ $f$ values obtained by sampling at transformed Halton points. Notice how a two-dimensional Halton generator is used to construct a Gaussian true measure, which is applied alongside the \texttt{my\_Keister} function to instantiate a customized, QMCPy-compatible integrand for this problem.
\begin{lstlisting}[style=Python]
>>> def my_Keister(t):
...     d = t.shape[1] # t is an (n x d) array
...     norm = np.sqrt((t**2).sum(1))
...     out = np.pi**(d/2)*np.cos(norm)
...     return out  # size n vector
...
>>> gauss = qp.Gaussian(qp.Halton(2), covariance=1/2)
>>> keister = qp.CustomFun(true_measure=gauss, g=my_Keister)
>>> x = keister.discrete_distrib.gen_samples(1000) 
>>> y = keister.f(x)
>>> y.mean()
1.809055768468628
\end{lstlisting}
We have no indication yet of how accurate our approximation is.  That topic is treated in the next section.  
Fig.~\ref{fig:ikc} visualizes sampling on the original integrand, $g$, and sampling on the transformed integrand, $f$. 

Another way to approximate the Keister integral in \eqref{eq:KeisterIntegral} is to write it as an integral with respect to the Lebesgue measure:
\begin{align*} 
	\mu 
	& = \int_{\mathbb{R}^d} \underbrace{\cos(\lVert \boldsymbol{t} \rVert) \exp(-\boldsymbol{t}^T \boldsymbol{t}) }_{g(\boldsymbol{t})} \, \underbrace{1}_{\lambda(\boldsymbol{t})} \,\D \boldsymbol{t} \\
	& = \int_{[0,1]^d} \underbrace{\cos(\lVert \boldsymbol{\Psi}(\boldsymbol{x}) \rVert) \exp(-\boldsymbol{\Psi}^T\!\!(\boldsymbol{x}) \boldsymbol{\Psi}(\boldsymbol{x})) \lvert \boldsymbol{\Psi}'(\boldsymbol{x})\rvert  }_{f(\boldsymbol{x})} \, \D \boldsymbol{x},
\end{align*}
where $\boldsymbol{\Psi}$ is any transformation from $[0,1]^d$ to $\mathbb{R}^d$. Now $\lambda$ is \emph{not} a PDF.  QMCPy can perform the cubature this way as well.
\begin{lstlisting}[style=Python]
>>> def my_L_Keister(t):
...     norm_sq = (t**2).sum(1)
...     out = np.cos(np.sqrt(norm_sq))*np.exp(-norm_sq)
...     return out
...
>>> lebesgue_gauss = qp.Lebesgue(qp.Gaussian(qp.Halton(2)))
>>> keister = qp.CustomFun(lebesgue_gauss, my_L_Keister)
>>> x = keister.discrete_distrib.gen_samples(1000)
>>> y = keister.f(x)
>>> y.mean()
1.8056340581961572
\end{lstlisting}
The $\boldsymbol{\Psi}$ chosen when transforming uniform sequences on the unit cube to fill $\mathbb{R}^d$ is  given by \eqref{eq:exampleVarTransGauss} with $\mathsf{A} = \mathsf{I}$.

In the examples above, one must input the correct $g$ into \texttt{CustomFun} along with the correct \texttt{TrueMeasure} $\lambda$ to define the integration problem. The \texttt{Keister} integrand included in the QMCPy library takes a more flexible approach to defining the integration problem $\mu$ in \eqref{eq:KeisterIntegral}. Selecting a different \texttt{sampler} $\boldsymbol{\Psi}$ performs  \emph{importance sampling}, which leaves $\mu$ unchanged.  
\begin{lstlisting}[style=Python]
>>> # default transform
>>> keister = qp.Keister(qp.Halton(2))
>>> x = keister.discrete_distrib.gen_samples(1e4)
>>> keister.f(x).mean()
1.8082377673556123
>>> # custom transform for importance sampling 
>>> keister = qp.Keister(sampler=qp.Gaussian(qp.Halton(2)))
>>> x = keister.discrete_distrib.gen_samples(1e4)
>>> keister.f(x).mean()
1.8080555069060817
\end{lstlisting}

In the first case above, the $\lambda$ in \eqref{eq:transVar} corresponds to the Gaussian density  with mean zero and variance $1/2$ by default, and the corresponding variable transformation, $\boldsymbol{\Psi}$, is chosen to make $\lambda\bigl(\boldsymbol{\Psi}(\boldsymbol{x})\bigr) \lvert \boldsymbol{\Psi}'(\boldsymbol{x})\rvert  = 1$ and $f(\boldsymbol{x}) = g\left(\boldsymbol{\Psi}(\boldsymbol{x}) \right)$.  In the second case,  we choose an importance sampling density $\lambda_{\textup{IS}}$, corresponding  to standard  Gaussian, and the variable transformation $\boldsymbol{\Psi}_{\textup{IS}}$ makes $\lambda_\textup{IS}\bigl(\boldsymbol{\Psi}_{\textup{IS}}(\boldsymbol{x})\bigr) \lvert\boldsymbol{\Psi}'_{\textup{IS}}(\boldsymbol{x})\rvert  = 1$.  Then
\begin{align}
	\nonumber 
	\mu &= \int_\mathcal{T} g(\boldsymbol{t}) \, \lambda(\boldsymbol{t}) \, \D \boldsymbol{t}  = \int_\mathcal{T} g(\boldsymbol{t}) \, \frac{\lambda(\boldsymbol{t})}{\lambda_\textup{IS}(\boldsymbol{t}) } \, \lambda_\textup{IS}(\boldsymbol{t}) \D \boldsymbol{t}  \\ 
	\nonumber
	& =  \int_{[0,1]^d} g\bigl(\boldsymbol{\Psi}_\textup{IS}(\boldsymbol{x})\bigr) \, \frac{\lambda\bigl(\boldsymbol{\Psi}_\textup{IS}(\boldsymbol{x})\bigr)} {\lambda_\textup{IS}\bigl(\boldsymbol{\Psi}_\textup{IS}(\boldsymbol{x})\bigr) } \lambda_\textup{IS}\bigl(\boldsymbol{\Psi}_\textup{IS}(\boldsymbol{x})\bigr) \,\lvert\boldsymbol{\Psi}'_{\textup{IS}}(\boldsymbol{x})\rvert  \, \D \boldsymbol{x} \\
	\nonumber
	& =  \int_{[0,1]^d} f_\textup{IS}(\boldsymbol{x}) \, \D \boldsymbol{x}  \\
	\label{eq:transVarImp}
	& \qquad \qquad \text{where } f_\textup{IS}(\boldsymbol{x})  = g\bigl(\boldsymbol{\Psi}_\textup{IS}(\boldsymbol{x})\bigr)  \,  \frac{\lambda\bigl(\boldsymbol{\Psi}_\textup{IS}(\boldsymbol{x})\bigr)} {\lambda_\textup{IS}\bigl(\boldsymbol{\Psi}_\textup{IS}(\boldsymbol{x})\bigr) }.
\end{align}
Because LD samples mimic $\mathcal{U}[0,1]^d$, choosing a different \texttt{sampler} is equivalent to choosing a different variable transform.

\section{Stopping Criteria} \label{sec:stopping_crit}

The \texttt{StoppingCriterion} object determines the number of samples $n$ that are required for the sample mean approximation $\widehat{\mu}$ to be within error tolerance $\varepsilon$ of the true mean $\mu$.  Several QMC stopping criteria have been implemented in QMCPy, including replications, stopping criteria that track the decay of the Fourier complex exponential or Walsh coefficients of the integrand \cite{HicJim16a,HicEtal17a,JimHic16a}, and stopping criteria based on
Bayesian credible intervals \cite{RatHic19a,JagHic22a}. 

The \texttt{CubQMCSobolG} stopping criterion used in the example below assumes the Walsh-Fourier coefficients of the integrand are absolutely convergent. The algorithm iteratively doubles the number of samples used in the integration and estimates the error using the decay of the Walsh coefficients \cite{HicJim16a}. When the estimated error is below the user-specified tolerance, it finishes the computation and returns the estimated integral.

Let us return to the Keister example from the previous section.  After setting up  a default \texttt{Keister} instance via a Sobol' \texttt{DiscreteDistribution}, we choose a \texttt{StoppingCriterion} object that matches the \texttt{DiscreteDistribution} and input our desired tolerance.  Calling the  \texttt{integrate} method returns the approximate integral plus some useful information about the computation.
\begin{lstlisting}[style=Python]
>>> keister = qp.Keister(qp.Sobol(2))
>>> stopping = qp.CubQMCSobolG(keister, abs_tol=1e-3)
>>> solution_qmc,data_qmc = stopping.integrate()
>>> data_qmc # equivalent to print(data_qmc)
LDTransformData (AccumulateData Object)
    solution        1.808
    error_bound     6.06e-04
    n_total         2^(13)
    time_integrate  0.008
CubQMCSobolG (StoppingCriterion Object)
    abs_tol         0.001
    rel_tol         0
    n_init          2^(10)
    n_max           2^(35)
Keister (Integrand Object)
Gaussian (TrueMeasure Object)
    mean            0
    covariance      2^(-1)
    decomp_type     PCA
Sobol (DiscreteDistribution Object)
    d               2^(1)
    dvec            [0 1]
    randomize       LMS_DS
    graycode        0
    entropy         326942311248945520670220938885737472885
    spawn_key       ()
\end{lstlisting}
The second output of the stopping criterion provides helpful diagnostic information.  This computation requires $n=2^{13}$ Sobol' points and $0.008$ seconds to complete.  The error bound is $0.000606$, which falls below the absolute tolerance.

\begin{figure}[t]
	\includegraphics[height=6cm]{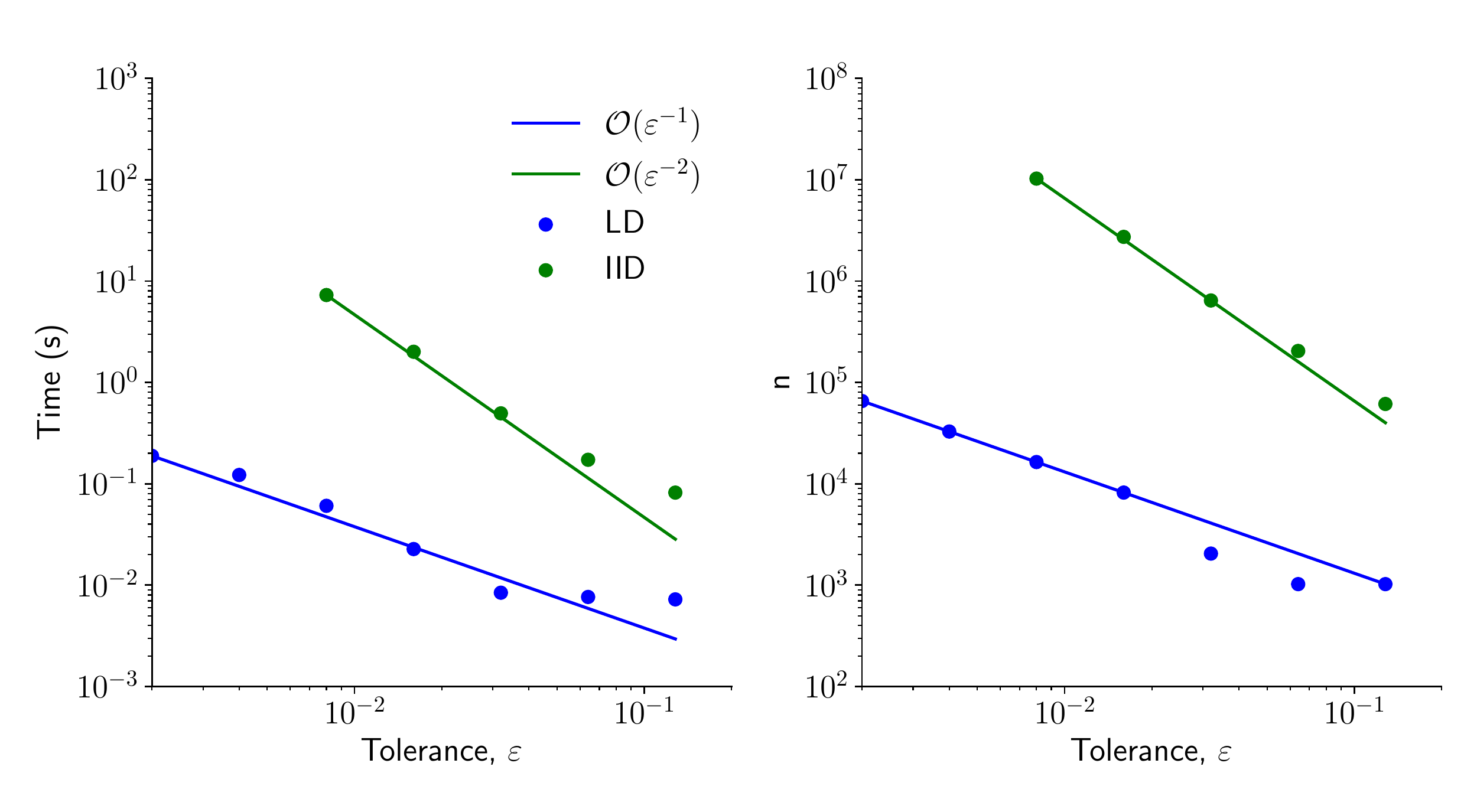}
	\caption{Comparison of run times and sample sizes for computing the $5$-dimensional Keister integral \eqref{eq:KeisterIntegral} using IID and LD lattice sequences for a variety of absolute error tolerances. The respective stopping criteria are  \texttt{qp.CubMCG} \cite{HicEtal14a} and  \texttt{qp.CubQMCLatticeG} \cite{JimHic16a}. The LD sequences provide the desired answer much more efficiently.}
	\label{fig:sc_comp}
\end{figure}

QMC, which uses LD sequences, is touted as providing substantially greater computational efficiency compared to IID MC.
Fig.~\ref{fig:sc_comp} compares the time and sample sizes needed to compute the $5$-dimensional Keister integral \eqref{eq:KeisterIntegral} using IID sequences and LD lattice sequences. Consistent with what is stated in Section \ref{sec:intro}, the error of IID MC is $\mathcal{O}(n^{-1/2})$, which means that the time and sample size to obtain an absolute error tolerance of $\varepsilon$ is $\mathcal{O}(\varepsilon^{-2})$.  By contrast, the  error of QMC using LD sequences is $\mathcal{O}(n^{-1+\epsilon})$, which implies $\mathcal{O}(\varepsilon^{-1-\epsilon})$ times and sample sizes.  We see that QMC methods often require orders of magnitude fewer samples than MC methods to achieve the same error tolerance.

For another illustration of QMC cubature, we turn to pricing an Asian arithmetic mean call option. The (continuous-time) Asian option is defined in terms of the average of the stock price, which is written in terms of an integral. The payoff of this option is the positive difference between the strike price, $K$, averaged over the time horizon: 
$$
\text{payoff}(\boldsymbol{S}) = \max\left(\frac{1}{2d}\sum_{j=1}^d (S_{j-1}+S_j)-K, \; 0\right), \quad \boldsymbol{S} = (S_0, \ldots, S_d).
$$
Here $S_j$ denotes the asset price at time $\tau j/d$, and a trapezoidal rule is used for discrete approximation of the integral in time that defines the average. 
The trapezoidal rule is a more accurate approximation to the integral than a rectangle rule. 
A basic model for asset prices is a geometric Brownian motion, 
\[
S_j(\boldsymbol{T}) = S_0 \exp((r - \sigma^2/2) \tau j/d + \sigma T_j),   \;  j = 1, \ldots, d, \; \boldsymbol{T} = (T_1, \ldots, T_d)\sim \mathcal{N}(\boldsymbol{0},\Sigma),
\]
where $\Sigma$ is defined in \eqref{eq:BMcov}, $r$ is the interest rate, $\sigma$ is the volatility, and $S_0$ is the initial asset price.  The fair price of the option is then the expected value of the discounted payoff, namely,
\begin{equation*}
	\text{price} = \mu = \mathbb{E}[g(\boldsymbol{T})], \quad \text{where } g(\boldsymbol{t}) = \text{payoff}\bigl(\boldsymbol{S}(\boldsymbol{t}) \bigr) \exp(-r \tau).
\end{equation*}

The following code utilizes QMCPy's Asian option \texttt{Integrand} object to approximate the value of an Asian call option for a particular choice of the parameters.
\begin{lstlisting}[style=Python]
>>> payoff = qp.AsianOption(
...     qp.Sobol(52), # weekly monitoring
...     start_price = 100,
...     strike_price = 120,
...     volatility = 0.5,
...     interest_rate = 0,
...     t_final = 1,
...     call_put = "call")
>>> qmc_stop_crit = qp.CubQMCSobolG(payoff, abs_tol=0.001)
>>> price,data = qmc_stop_crit.integrate()
>>> print("Option price = $%.3f using %.3f seconds, %.2e samples"
...     %(price, data.time_integrate, data.n_total))
Option price = $5.194 using 0.587 seconds, 1.31e+05 samples
\end{lstlisting}
Because this \texttt{Integrand} object has the built-in Brownian motion \texttt{TrueMeasure}, one only need provide the LD sampler.

Out of the money option price calculations can be sped up by adding an upward drift to the Brownian motion.  The upward drift produces more in the money paths and also reduces the variation or variance of the final integrand, $f$.  This is a form of importance sampling.  Using a Brownian motion without drift we get
\begin{lstlisting}[style=Python]
>>> payoff = qp.AsianOption(qp.Sobol(52),
...     start_price=100, strike_price=200)
>>> qmc_stopper = qp.CubQMCSobolG(payoff, abs_tol=0.001)
>>> price,data = qmc_stopper.integrate()
>>> print("Option price = $%.4f using %.3f seconds, n = %.2e"
...     %(price, data.time_integrate, data.n_total))
Option price = $0.1757 using 0.583 seconds, n = 1.31e+05
\end{lstlisting}
Adding the upward drift gives us the answer faster:
\begin{lstlisting}[style=Python]
>>> payoff = qp.AsianOption(
...     sampler = qp.BrownianMotion(qp.Sobol(52), drift=1),
...     start_price=100, strike_price=200)
>>> qmc_stopper = qp.CubQMCSobolG(payoff, abs_tol=0.001)
>>> price,data_drift = qmc_stopper.integrate()
>>> print("Option price = $%.4f using %.3f seconds, n = %.2e"
...     %(price, data_drift.time_integrate, data_drift.n_total))
Option price = $0.1754 using 0.085 seconds, n = 1.64e+04
>>> print("Using drift required %.0f%% the time, %.0f%% the n"
...     %(100*data_drift.time_integrate / data.time_integrate,
...       100*data_drift.n_total / data.n_total))
Using drift required 15% the time, 12% the n
\end{lstlisting}
The choice of a good drift is an art.  

The improvement in time is less than that in $n$ because the integrand is more expensive to compute when the drift is employed.  Referring to \eqref{eq:transVarImp}, in the case of no drift, the $\lambda$ corresponds to the density for the discrete Brownian motion, and the variable transformation $\boldsymbol{\Psi}$ is chosen so that $f(\boldsymbol{x}) = g\left(\boldsymbol{\Psi}(\boldsymbol{x}) \right)$.  However, in the case of a drift, the integrand becomes  $f_\textup{IS}(\boldsymbol{x})  = g\bigl(\boldsymbol{\Psi}_\textup{IS}(\boldsymbol{x})\bigr)  \lambda\bigl(\boldsymbol{\Psi}_\textup{IS}(\boldsymbol{x})\bigr)/\lambda_\textup{IS}\bigl(\boldsymbol{\Psi}_\textup{IS}(\boldsymbol{x})\bigr)$, which requires more computation time per integrand value. 

\section{Under the Hood}

In this section, we look at the inner workings of QMCPy and point out features we hope will benefit the community of QMC researchers and practitioners. We also highlight important nuances of QMC methods and how QMCPy addresses these challenges. For details, readers should refer to the QMCPy documentation \cite{QMCPyDocs}.

\subsection{LD Sequences}

LD sequences are the backbone of QMC methods. QMCPy provides generators that combine research from across the QMC community to enable advanced features and customization options.

Two popular LD sequences are integration lattices and digital nets which we previously outlined in Table \ref{tab:GroupProp}. These LD generators are comprised of two parts: the static generating vectors $\boldsymbol{Z}_1,\boldsymbol{Z}_2,\boldsymbol{Z}_4, \ldots \in [0,1)^d$ and the callable generator function. By default, QMCPy provides a number of high-quality generating vectors for users to choose from. For instance, the default ordinary lattice vector was constructed by Cools, Kuo, and Nuyens \cite{doi:10.1137/06065074X} using component-by-component search, is extensible, has order-2 weights, and supports up to 3600 dimensions and $2^{20}$ samples. 
However, users who require more samples but fewer dimensions may switch to a generating vector constructed using LatNet Builder \cite{LatNet,LEcEtal22a} to support 750 dimensions and $2^{24}$ samples. Moreover, the \texttt{qp.Lattice} and \texttt{qp.DigitalNet} objects allow users to input their own generating vectors to produce highly customized sequences. To find such vectors, we recommend using LatNet Builder's construction routines as the results can be easily parsed into a QMCPy-compatible format. 

Along with the selection of a generating vector, QMCPy's low discrepancy sequence routines expose a number of other customization parameters. For instance, the lattice generator extends the Magic Point Shop \cite{Nuy17a} to support either linear or natural ordering. Digital sequences permit either standard or Gray code ordering and may be randomized via a digital shift optionally combined with a linear scrambling~\cite{Mat98}. Halton sequences may be randomized via the routines of either Owen~\cite{Owe20a} or Hofert and Lemieux \cite{QRNG2020}. 

\subsection{A Word of Caution When Using LD Sequences}

Although QMCPy's \texttt{DiscreteDistribution}s have many of the same parameters and methods, users should be careful when swapping IID sequences with LD sequences. While IID node-sets have no preferred sample size, LD sequences often require special sampling ranges to ensure optimal discrepancy. As mentioned  earlier, base-2 digital nets and extensible integration lattices show better evenness for sample sizes that are powers of 2. On the other hand, the preferred sample sizes for $d$-dimensional Halton sequences are $n = \prod_{j=1}^d p^{m_j}_j$ where $p_j$ is the $j^{\text{th}}$ prime number and $m_j \in \mathbb{N}_0$ for $j=1,\dots,d$. Due to the infrequency of such values, Halton sequences are often regarded as not having a preferred sample size.  

Users may also run into trouble when trying to generate too many points. Since QMCPy's generators construct sequences in 32-bit precision, generating greater than~$2^{32}$ consecutive samples will cause the sequence to repeat. In the future, we plan to expand our generators to support optional 64-bit precision at the cost of greater computational overhead.

Another subtlety arises when transforming LD sequences to mimic different distributions. As mentioned earlier, unrandomized lattice and digital sequences include the origin, making transformations such as \eqref{eq:exampleVarTransGauss} produce infinite values.  

Some popular implementations of LD sequences drop the first point, which is the origin in the absence of randomization.  The rationale is to avoid the transformation of the origin to infinity when mimicking a Gaussian or other distribution with an infinite sample space.  Unfortunately, dropping the first point destroys some nice properties of the first $n = 2^m$ points of LD sequences, which can degrade the order of convergence for QMC cubature. A careful  discussion of this matter is given by~\cite{owen2020dropping}.

\subsection{Transformations}

The transformation $\boldsymbol{\Psi}$  connects a \texttt{DiscreteDistribution} and \texttt{TrueMeasure}. So far, we have assumed the \texttt{DiscreteDistribution} mimics a $\mathcal{U}[0,1]^d$ distribution with PDF $\varrho(\boldsymbol{x})=1$. However, it may be advantageous to utilize a \texttt{DiscreteDistribution} that mimics a different distribution. 

Suppose we have a \texttt{DiscreteDistribution} mimicking  density $\varrho$ supported on $\mathcal{X}$.  Then using the variable transformation $\boldsymbol{\Psi}$, 
\begin{equation*}
	\mu = \int_\mathcal{T} g(\boldsymbol{t}) \, \lambda(\boldsymbol{t}) \, \D \boldsymbol{t}  =  \int_{\mathcal{X}}f(\boldsymbol{x}) \, \varrho(\boldsymbol{x}) \D \boldsymbol{x} \quad \text{for }
   f(\boldsymbol{x})  = g\bigl(\boldsymbol{\Psi}(\boldsymbol{x})\bigr)  \, \frac{\lambda\bigl(\boldsymbol{\Psi}(\boldsymbol{x})\bigr)}{\varrho(\boldsymbol{x})} \, \lvert \boldsymbol{\Psi}'(\boldsymbol{x})\rvert,
\end{equation*}
which generalizes \eqref{eq:transVar}.
QMCPy also includes support for successive changes of measures so users may build complex variable transformations in an intuitive manner. Suppose that the variable transformation is a composition of several transformations: $\boldsymbol{\Psi}=\widehat{\boldsymbol{\Psi}}_L = \boldsymbol{\Psi}_L \circ \boldsymbol{\Psi}_{L-1} \circ \dots \circ \boldsymbol{\Psi}_1$ as in \eqref{eq:transVarImp}. Here, $\boldsymbol{\Psi}_l:\mathcal{X}_{l-1} \to \mathcal{X}_l$,  $\mathcal{X}_0=\mathcal{X}$, and $\mathcal{X}_L=\mathcal{T}$ so that the transformations are compatible with the \texttt{DiscreteDistribution} and \texttt{TrueMeasure}. Let $\widehat{\boldsymbol{\Psi}}_l=\boldsymbol{\Psi}_l \circ \boldsymbol{\Psi}_{l-1} \circ \dots \circ \boldsymbol{\Psi}_1$ denote the composition of the first $l$ transforms and assume that $\widehat{\boldsymbol{\Psi}}_0(\boldsymbol{x})=\boldsymbol{x}$, the identity transform. Then we may write $\mu =  \int_{\mathcal{X}}f(\boldsymbol{x}) \, \varrho(\boldsymbol{x}) \D \boldsymbol{x}$ for 
\begin{equation*}
    f(\boldsymbol{x}) 
 = g\bigl(\widehat{\boldsymbol{\Psi}}_L(\boldsymbol{x})\bigr)\frac{\lambda\bigl(\widehat{\boldsymbol{\Psi}}_L(\boldsymbol{x})\bigr)}{\varrho(\boldsymbol{x})}\prod_{l=1}^L \bigl\lvert \boldsymbol{\Psi}_l'\bigl(\widehat{\boldsymbol{\Psi}}_{l-1}(\boldsymbol{x})\bigr) \bigr\rvert .
\end{equation*}
It is often the case that $\boldsymbol{\Psi}_l$ is chosen such that $\boldsymbol{\Psi}_l(\boldsymbol{X})$ is stochastically equivalent to a random variable with density $\lambda_l$ on sample space $\mathcal{X}_l$ when $\boldsymbol{X}$ is a random variable with density $\varrho_l$ on sample space $\mathcal{X}_{l-1}$.  This implies  $\varrho_l(\boldsymbol{x}) = \lambda_l(\boldsymbol{\Psi}_l(\boldsymbol{x}))\lvert  \boldsymbol{\Psi}'_l (\boldsymbol{x}) \rvert$ so that
\begin{equation*}
f(\boldsymbol{x}) 
 = g\bigl(\widehat{\boldsymbol{\Psi}}_L(\boldsymbol{x})\bigr)\frac{\lambda\bigl(\widehat{\boldsymbol{\Psi}}_L(\boldsymbol{x})\bigr)}{\varrho(\boldsymbol{x})}\prod_{l=1}^L \frac{\varrho_l(\widehat{\boldsymbol{\Psi}}_{l-1}(\boldsymbol{x}))}{\lambda_l(\widehat{\boldsymbol{\Psi}}_l(\boldsymbol{x}))} .
\end{equation*}

For an example, we return to the Keister integral \eqref{eq:KeisterIntegral}. The following code constructs three \texttt{Keister} instances: one without importance sampling, one importance sampled by a Gaussian distribution, and one importance sampled by the composition of a Gaussian distribution with a Kumaraswamy distribution \cite{Kumaraswamy}. All \texttt{Integrand}s use a Sobol' \texttt{DiscreteDistribution}, making $\varrho(\boldsymbol{x})=1$ and $\mathcal{X} = [0,1]^d$. The \texttt{TrueMeasure} is $\mathcal{N}(\boldsymbol{0},\mathsf{I}/2)$ making $\lambda(\boldsymbol{t}) = \pi^{-d/2} \exp(-\boldsymbol{t}^T \boldsymbol{t})$ and $\mathcal{T} = \mathbb{R}^d$. 

The table below displays the variable transformations and the measures for these three cases.  In all cases $\varrho_1(\boldsymbol{x}) = \cdots = \varrho_L(\boldsymbol{x}) = 1$ because the $\boldsymbol{\Psi}_l$ utilize inverse cumulative distributions.

\[\begin{array}{ccccccc}
	\texttt{Integrand} & L & \lambda_1 & \boldsymbol{\Psi}_1 & \lambda_2 & \boldsymbol{\Psi}_2 & f \tabularnewline%
	\hline
	\texttt{K} & 1 & \mathcal{N}(\boldsymbol{0}, \mathsf{I}/2) & \eqref{eq:exampleVarTransGauss}& && g(\boldsymbol{\Psi}_1(\cdot)) \tabularnewline [1ex]
	\texttt{K\_gauss} & 1 & \mathcal{N}(\boldsymbol{0},3 \mathsf{I}/4) & \eqref{eq:exampleVarTransGauss} & && 
	\displaystyle g(\boldsymbol{\Psi}_1(\cdot)) \frac{\lambda(\boldsymbol{\Psi}_1(\cdot))}{\lambda_1(\boldsymbol{\Psi}_1(\cdot))} \tabularnewline [2.5ex]
	\texttt{K\_gauss\_kuma} & 2 & \textup{Kum} & \boldsymbol{F}^{-1}_\textup{Kum} & \mathcal{N}(\boldsymbol{0},\mathsf{I}) &  \eqref{eq:exampleVarTransGauss} &
	\displaystyle \frac{g(\boldsymbol{\Psi}_2(\boldsymbol{\Psi}_1(\cdot))) \lambda(\boldsymbol{\Psi}_2(\boldsymbol{\Psi}_1(\cdot))) }{\lambda_1(\boldsymbol{\Psi}_1(\cdot)) \lambda_2(\boldsymbol{\Psi}_2(\boldsymbol{\Psi}_1(\cdot)))} 
   \\ \hline 
\end{array}
\]

Here $\textup{Kum}$ denotes the multivariate Kumaraswamy distribution with independent marginals, and $ \boldsymbol{F}^{-1}_{\textup{Kum}}$ denotes the element-wise inverse cumulative distribution function.  The  code below evaluates the Keister integral \eqref{eq:KeisterIntegral} for $d=1$ and error tolerance $\varepsilon = 5 \times 10^{-8}$.  The timings for each of these different integrands are displayed.

\begin{lstlisting}[style=Python]
>>> sobol = qp.Sobol(1)
>>> K = qp.Keister(sobol) # keister 0
>>> K_gauss = qp.Keister( # keister 1
...     qp.Gaussian(sobol, covariance=.75))
>>> K_gauss_kuma = qp.Keister( # keister 2
...     qp.Gaussian(
...         qp.Kumaraswamy(sobol, a=.8, b=.8)))
>>> for i,keister in enumerate([K, K_gauss, K_gauss_kuma]):
...     stopper = qp.CubQMCSobolG(keister, abs_tol=5e-8)
...     sol,data = stopper.integrate()
...     print("keister %d integration time = %.3f seconds"
...         %(i, data.time_integrate))
keister 0 integration time = 0.815 seconds
keister 1 integration time = 0.338 seconds
keister 2 integration time = 0.040 seconds
\end{lstlisting}

\begin{figure}[t]
    \centering
	\includegraphics[width=.75\textwidth]{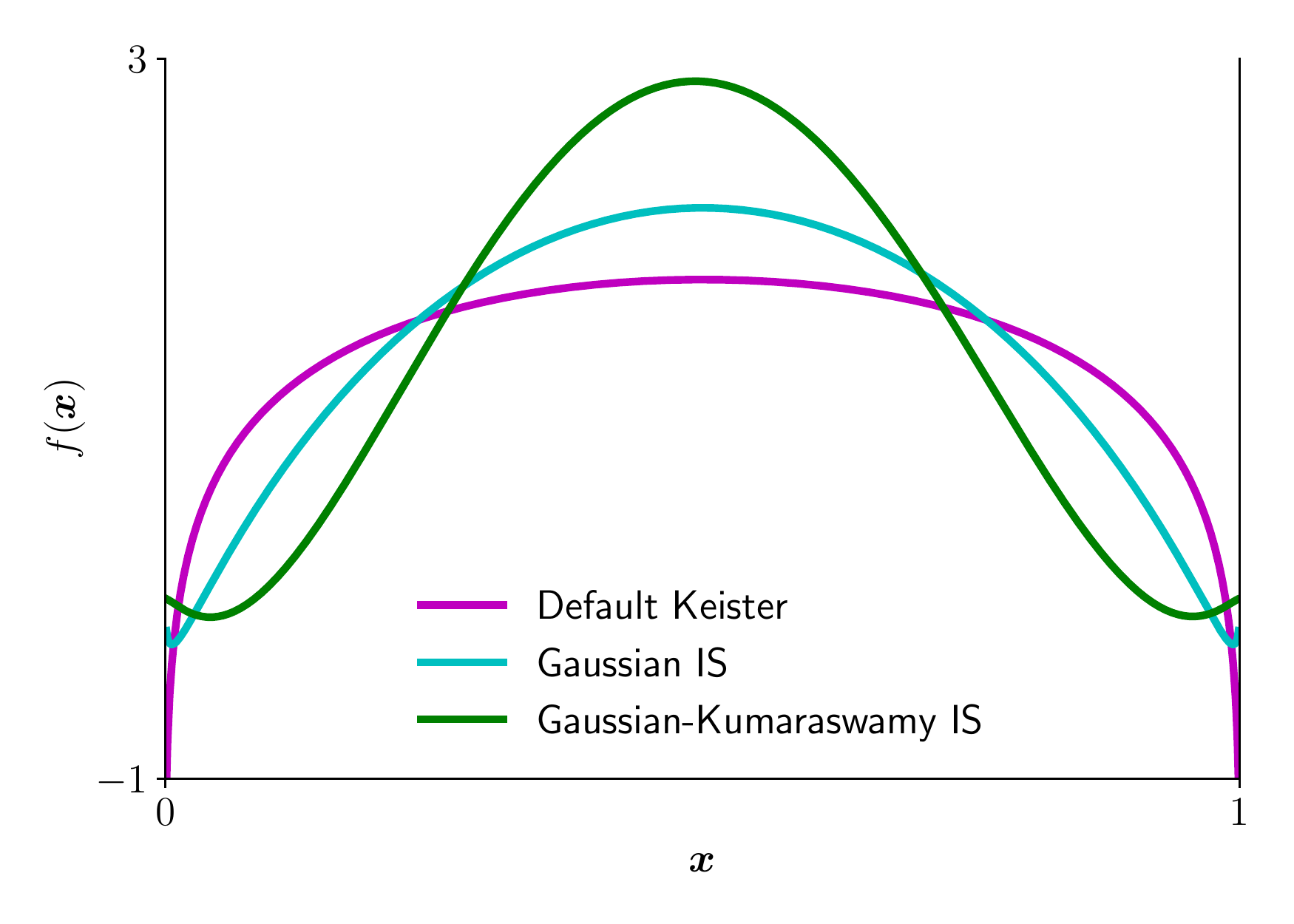}
	\caption{Keister functions with and without importance sampling (\texttt{IS}). Note that the Keister functions using importance sampling are generally less variable and therefore easier to integrate, as evidenced by the faster integration times.} \label{fig:mIS}
\end{figure}

Successful importance sampling makes the transformed integrand, $f$, more flat. The shorter cubature times correspond to flatter integrands, as  illustrated in Fig.~\ref{fig:mIS}. The above example uses $d=1$ to facilitate the plot in 
Fig.~\ref{fig:mIS}; however, the same example works for arbitrary dimensions.


\section{Further Work} \label{sec:further}

QMCPy is ripe for growth and development in several areas.  We hope that the QMC community will join us in making this a reality.

Multi-level (quasi-)Monte Carlo (ML(Q)MC) methods make possible the computation of expectations of functionals of stochastic differential equations and partial differential equations with random coefficients.  Such use cases appear in quantitative finance and geophysical applications.  QMCPy's ML(Q)MC's capability is rudimentary, but under active development.

We hope to add a greater variety of use cases and are engaging collaborators to help.  Sobol' indices, partial differential equations with random coefficients, expected improvement measures for Bayesian optimization, and multivariate probabilities are some of those on our radar.

Recently, several QMC experts have focused on developing LD generators for Python. 
Well-established packages such as SciPy \cite{SCIPY} and PyTorch \cite{PyTorch} are have developed QMC modules that support numerous LD sequences and related functionalities. We plan to integrate the routines as optional backends for QMCPy's LD generators. Creating ties to these other packages will allow users to call their preferred generators from within the QMCPy framework.  Moreover, as features in QMCPy become more common and prove their value, we will try to incorporate them into SciPy and other popular, general-purpose packages.

We also plan to expand our library of digital net generating matrices. We wish to incorporate interlaced digital nets, polynomial lattices, and Niederreiter sequences, among others. By including high-quality defaults in QMCPy, we hope to make these sequences more readily available to the public. 

Our \texttt{DiscreteDistrution} places equal weights on each support point, $\boldsymbol{X}_i$.  In the future, we might generalize this to unequal weights.

QMCPy already includes importance sampling, but the choice of sampling distribution must be chosen a priori.  We would like to see an automatic, adaptive choice following the developments of \cite{AsmGly2007a,approx_zero_variance_simulation,OweZho00a}.

Control variates can be useful for QMC as well as for IID MC \cite{HicEtal03}.  These should be incorporated into QMCPy in a seamless way.

We close with an invitation.  Try QMCPy.  If you find bugs or missing features, please submit an issue to https://github.com/QMCSoftware/QMCSoftware/issues.  
If you wish to add your great algorithm or use case, please submit a pull request to our GitHub repository at https://github.com/QMCSoftware/QMCSoftware/pulls. 
We hope that the community will embrace QMCPy.

\begin{acknowledgement}
The authors would like to thank the organizers for a wonderful MCQMC 2020. 
We also thank the referees for their many helpful suggestions.  This work is supported in part by SigOpt and National Science Foundation grant DMS-1522687.
\end{acknowledgement}


\end{document}